\documentclass[12pt]{article}

\usepackage{colortbl,arydshln}
\usepackage[normalem]{ulem}
\usepackage{dsfont}
\usepackage{longtable}
\usepackage[hang,small,bf]{caption}
\usepackage{multirow,multicol}
\usepackage{graphicx}
\usepackage{epstopdf}
\usepackage{textcomp}
\usepackage[utf8]{inputenc}

\usepackage{amsmath,amssymb,latexsym,float,amsthm}
\usepackage{array}
\usepackage{graphicx,color}
\usepackage{lipsum}
\usepackage[numbers,sort&compress]{natbib}
\usepackage[english]{babel}
\usepackage{geometry}
\usepackage{bbm}

\DeclareGraphicsExtensions{.pdf,.eps,.png,.jpeg}

\newcommand{\bbone}{\ensuremath{\mathbbm{1}}}
\newcommand{\QI}{\psi}
\newcommand{\PC}{\theta}
\newcommand{\Nc}{L}
\newcommand{\nc}{\ell}
\newcommand{\Varnc}{Z}
\newcommand{\Nbtab}{M}
\newcommand{\nbtab}{m}

\newcommand{\Nbind}{I}
\newcommand{\nbind}{i}
\newcommand{\Nbvar}{K}
\newcommand{\nbvar}{k}
\newcommand{\Nbcol}{J}
\newcommand{\nbcol}{j}
\newcommand{\Nbsim}{T}
\newcommand{\nbsim}{t}
\newcommand{\Nbdim}{S}
\newcommand{\nbdim}{s}
\newcommand{\bfD}{{\textbf{D}}}
\newcommand{\bfX}{{\textbf{X}}}
\newcommand{\bfx}{{\textbf{x}}}

\newcommand{\bfV}{{\textbf{V}}}
\newcommand{\bfU}{{\textbf{U}}}
\newcommand{\bfW}{{\textbf{W}}}
\newcommand{\bfM}{{\textbf{M}}}
\newcommand{\bfZ}{{\textbf{Z}}}
\newcommand{\bfz}{{\textbf{z}}}
\def\keywords{\vspace{.5em}
{\textit{Keywords}:\,\relax%
}}

\begin{document}

\begin{center}
{\Large{MIMCA: Multiple imputation for categorical variables with multiple correspondence analysis}}
\end{center}
 \vglue-2cm
\begin{center}
  \scshape Vincent \textsc{Audigier}\footnote{Principal corresponding author}, Fran\c cois
  \textsc{Husson}\footnote{Corresponding author} and Julie \textsc{Josse}\footnotemark[2]
\renewcommand{\thefootnote}{\arabic{footnote}}\setcounter{footnote}{0}
\end{center}
\vglue0.3cm
\hglue0.02\linewidth\begin{minipage}{0.9\linewidth}
\begin{center}
Applied Mathematics Department,
Agrocampus Ouest, 65 rue de Saint-Brieuc,
F-35042 RENNES Cedex, France \\
\parbox[t]{0.45\linewidth}{\texttt{audigier@agrocampus-ouest.fr}
\texttt{husson@agrocampus-ouest.fr} \texttt{josse@agrocampus-ouest.fr}}
\end{center}
\end{minipage}
\begin{abstract}
We propose a multiple imputation method to deal with incomplete categorical data. This method imputes the missing entries using the principal components method dedicated to categorical data: multiple correspondence analysis (MCA). The uncertainty concerning the parameters of the imputation model is reflected using a non-parametric bootstrap. Multiple imputation using MCA (MIMCA) requires estimating a small number of parameters due to the dimensionality reduction property of MCA.  It allows the user to impute a large range of data sets. In particular, a high number of categories per variable, a high number of variables or a small the number of individuals are not an issue for MIMCA. Through a simulation study based on real data sets, the method is assessed and compared to the reference methods (multiple imputation using the loglinear model, multiple imputation by logistic regressions) as well to the latest works on the topic (multiple imputation by random forests or by the Dirichlet process mixture of products of multinomial distributions model). The proposed method shows good performances in terms of bias and coverage for an analysis model such as a main effects logistic regression model. In addition, MIMCA has the great advantage that it is substantially less time consuming on data sets of high dimensions than the other multiple imputation methods.
\end{abstract}
\keywords{missing values, categorical data, multiple imputation, multiple correspondence analysis, bootstrap}
\section{Introduction}
Data sets with categorical variables are ubiquitous in many fields such in social sciences, where surveys are conducted through multiple-choice questions. Whatever the field, missing values frequently occur and are a key problem in statistical practice since most of statistical methods cannot be applied directly on incomplete data. 

To deal with missing values one solution consists in adapting the statistical method so that it can be applied on an incomplete data set. For instance, the maximum likelihood (ML) estimators can be derived from incomplete data using an Expectation-Maximization (EM) algorithm \cite{Dempster77} and their standard error can be estimated using a Supplemented Expectation-Maximization algorithm \cite{Meng91}. The ML approach is suitable, but not always easy to establish \cite{Allison12}.

\begin{figure}[h!]
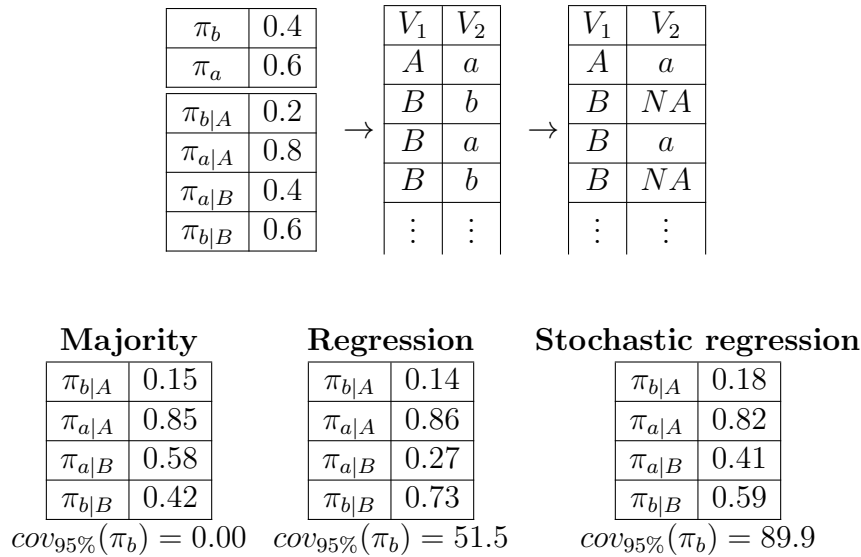

\begin{center}
$$\begin{array}{ccc}

\begin{array}{|c|c|}
\hline
\pi_b&0.4\\ \hline
\pi_a&0.6\\ \hline \hline
 \pi_{b|A}&  0.2 \\ \hline
 \pi_{a|A} & 0.8 \\ \hline
 \pi_{a|B}&  0.4 \\ \hline
 \pi_{b|B}&  0.6\\ \hline
\end{array}

&\rightarrow
\begin{array}{|c|c|}
\hline V_1 &V_2\\ \hline
A  &a\\ \hline
B  &b \\ \hline
 B  &a\\ \hline
  B  &b \\ \hline
 \vdots&\vdots
 \end{array}
 
 &\rightarrow
 \begin{array}{|c|c|}
\hline V_1 &V_2\\ \hline
A  &a\\ \hline
 B  &{NA} \\ \hline
 B  &a\\ \hline
  B  &{NA} \\ \hline
 \vdots&\vdots
 \end{array}
 
\end{array}$$

\vfill
$$\begin{array}{ccc}

\textbf{Majority}&\textbf{Regression}& \textbf{Stochastic regression}\\
\begin{array}{|c|c|}
\hline \pi_{b|A}& 0.15  \\ \hline
\pi_{a|A}& 0.85  \\ \hline
\pi_{a|B}& 0.58  \\ \hline
\pi_{b|B}& 0.42   \\ \hline
 \end{array}
&

\begin{array}{|c|c|}
\hline  \pi_{b|A}& 0.14  \\ \hline
\pi_{a|A}& 0.86  \\ \hline
\pi_{a|B}& 0.27  \\ \hline
\pi_{b|B}& 0.73  \\ \hline
 \end{array}
&

\begin{array}{|c|c|}
\hline  \pi_{b|A}& 0.18  \\ \hline
\pi_{a|A}& 0.82  \\ \hline
\pi_{a|B}& 0.41  \\ \hline
\pi_{b|B}& 0.59  \\ \hline
\end{array}
\\

cov_{95\%}(\pi_b)=0.00 &cov_{95\%}(\pi_b)=51.5 &cov_{95\%}(\pi_b)={89.9}
\end{array}$$
\caption{Illustration of three imputation methods for two categorical variables\label{fig1}: the top part described how the data are built (marginal and conditional proportions, associated complete data, incomplete data set generated where NA denotes a missing value) and the bottom part sums up the observed conditional proportions after an imputation by several methods (majority, regression, stochastic regression). The last line indicates the coverage for the confidence interval for the proportion of $b$ over 1000 simulations.}
\end{center}
\end{figure}

Another way consists in replacing missing values by plausible values according to an \textit{imputation model}. This is called \textit{single imputation}. Thus, the data set is complete and any statistical method can be applied on this one. Figure \ref{fig1} illustrates three simple single imputation methods.
The data set used contains 1000 individuals and two variables with two categories: $A$ and $B$ for the first variable, $a$ and $b$ for the second one. The data set is built so that 40\% of the individuals take the category $a$ and 60\% the category $b$. In addition, the variables are linked, that is to say, the probability to observe $a$ or $b$ on the second variable depends on the category taken on the first variable. Then, 30\% of missing values are generated completely at random on the second variable. A first method could be to impute according to the most taken category of the variable. In this case, all missing values are imputed by $a$. Consequently, marginal proportions are modified, as well as conditional proportions (see the bottom part of Figure \ref{fig1}).
This method is clearly not suitable. A more convenient solution consists in taking into account the relationship between the two variables, following the rationale of the imputation by regression for continuous data. To achieve this goal, the parameters of a logistic regression are estimated from the complete cases, providing fitted conditional proportions. Then, each individual is imputed according to the highest conditional proportion given the first variable. This method respects the conditional proportions better, but the relationship between variables is strengthened which is not satisfactory. In order to obtain an imputed data set with a structure as close as possible to the generated data set, a suitable single imputation method is to perform stochastic regression: instead of imputing according to the  the most likely category, the imputation is performed randomly according to the fitted probabilities. 
 
An imputation model used to perform single imputation has to be sufficiently complex compared to the statistical method desired (the \textit{analysis model}). For instance, if the aim is to apply a logistic regression from an incomplete data set, it requires using an imputation model taking into account the relationships between variables. Thus, a suitable single imputation method, such as the stochastic regression strategy,  leads to unbiased estimates of the parameters of the statistical method (see Figure \ref{fig1}). However, although the single imputation method respects the structure of the data, it still has the drawback that it leads to underestimate the variability of the estimators because the uncertainty on the imputed values is not taken into account in the estimate of the variability of the estimators. However, although the single imputation method respects the structure of the data, it still has the drawback that the uncertainty on the imputed values is not taken into account in the estimate of the variability of the estimators. Thus, this variability remains underestimated. For instance, in Figure \ref{fig1}, the level of the confidence interval of $\pi_b$, the proportion of $b$, is 89.9\% and does not reach the nominal rate of 95\%.

Multiple imputation (MI) \cite{Rubin87,Little02}  has been developped to avoid this issue. The principle of multiple imputation consists in creating $\Nbtab$ imputed data sets to reflect the uncertainty on imputed values. Then, the parameters of the statistical method, denoted $\QI$, are estimated from each imputed data set, leading to $\Nbtab$ sets of parameters $(\widehat\QI_\nbtab)_{1\leq\nbtab\leq \Nbtab}$. Lastly, these sets of parameters are pooled to provide a unique estimation for $\QI$ and for its associated variability using  Rubin's rules \cite{Rubin87}.

MI is based on the \textit{ignorability} assumption, that is to say ignoring the mechanism that generated missing values. This assumption  is equivalent to: first, the parameters that govern the missing data mechanism and the parameters of the analysis model are independent; then, missing values are generated \textit{at random}, that is to say, the probability that a missing value occurs on a cell is independent from the value of the cell itself. In practice, ignorability and value missing at random (MAR), are used interchangeably. This assumption is more plausible when the number of variables is high \cite{Schafer97,vanderPalm14}, but remains difficult to verify.

Thus, under the ignorability assumption, the main challenge in multiple imputation is to reflect the uncertainty of the imputed values by reflecting \textit{properly} \cite[p. 118-128]{Rubin87} the uncertainty on the parameters of the model used to perform imputation to get  imputed data sets yielding to valid statistical inferences. To do so, two classical approaches can be considered. The first one is the Bayesian approach: a prior distribution is assumed on the parameters $\PC$ of the imputation model, it is combined with the observed entries, providing a posterior distribution from which $\Nbtab$ sets of parameters $\left(\tilde{\PC}_{\nbtab}\right)_{1\leq\nbtab\leq\Nbtab}$ are drawn. Then, the incomplete data set is imputed $\Nbtab$ times using each set of parameters. The second one is a bootstrap approach: $\Nbtab$ samples with replacement are drawn leading to $\Nbtab$ incomplete data sets from which the parameters of the imputation model are obtained. The $\Nbtab$ sets of parameters $\left(\PC_{\nbtab}\right)_{1\leq\nbtab\leq\Nbtab}$ are then used to perform $\Nbtab$ imputations of the original incomplete data set.

In this paper, we detail in Section 2 the main available MI methods to deal with categorical data. Two general modelling strategies can be distinguished for imputing multivariate data: joint modelling (JM) \cite{Schafer97} and fully conditional specification (FCS)\cite{vanBuuren06}. JM is based on the assumption that the data can be described by a multivariate distribution. Concerning FCS, the multivariate distribution is not defined explicitly, but implicitly through the conditional distributions of each variable only. Among the presented methods, three are JM methods: MI using the loglinear model, MI using the latent class model and MI using the normal distribution; the two others are FCS strategies: the FCS using logistic regressions and FCS using random forests \cite{Breiman01}. In Section 3, a novel JM method based on a principal components method dedicated to categorical data, namely multiple correspondence analysis (MCA), is proposed. Principal components methods are commonly used to highlight the similarities between individuals and the relationships between variables, using a small number of principal components and loadings. MI based on this family of methods uses these similarities and these relationships to perform imputation, while using a restricted number of parameters. The performances of the imputation are very promising from continuous data \cite{Audigier15,Josse11} which motivates the consideration of a method for categorical data. 
 In Section 4, a simulation study based on real data sets, evaluates the novel method and compares its performances to other main multiple imputation methods. Lastly, conclusions about MI for categorical data and possible extensions for the novel method are detailled.

\section{Multiple imputation methods for categorical data\label{sec2}}
The imputation of categorical variables is rather complex. Indeed, contrary to continuous data, the variables follow a distribution on a discrete support defined by the combinations of categories observed for each individual. Because of the explosion of the number of combinations when the number of categories increases, the number of parameters defining the multivariate distribution could be extremely large. Consequently, defining an imputation model is not straightforward for categorical data. In this section we review the most popular approaches commonly used to deal with categorical data: JM using the loglinear model, JM using the latent class model, JM using the normal distribution and FCS using multinomial logistic regression or random forests.


Hereinafter, matrices and vectors will be in bold text, whereas sets of random variables or single random variables will not. Matrices will be in capital letters, whereas vectors will be in lower case letters. We denote $\bfX_{\Nbind \times\Nbvar}$ a data set with $\Nbind$ individuals and $\Nbvar$ variables. We note the observed part of $\bfX$ by $\bfX_{obs}$ and the missing part by $\bfX_{miss}$, so that $\bfX=\left(\bfX_{obs},\bfX_{miss}\right)$. Let $q_{\nbvar}$ denote the number of categories for the variable $\bfX_{\nbvar}$, $\Nbcol=\sum_{\nbvar=1}^{\Nbvar}q_{\nbvar}$ the total number of categories. We note $\mathbb{P}\left(X,\PC \right)$ the distribution of the variables $X=(X_1,\ldots,X_{\Nbvar})$, where $\PC$ is the corresponding set of parameters.

\subsection{Multiple imputation using a loglinear model}
The saturated loglinear model (or multinomial model) \cite{Agresti02} consists in assuming a multinomial distribution $\mathcal{M}\left(\PC,1\right)$ as joint distribution for $X$, where $\PC=\left(\PC_{x_1\ldots x_{\Nbvar}}\right)_{x_1\ldots x_{\Nbvar}}$ is a vector indicating the probability to observe each event $\left(X_1=x_1,\ldots, X_{\Nbvar}=x_{\Nbvar}\right)$. Performing MI with the loglinear model \cite{Schafer97} is often achieved by reflecting the variability of the imputation model's parameters with a Bayesian approach. More precisely, a Bayesian treatment of this model can be specified as follows:
\begin{eqnarray}
X\vert \PC \sim \mathcal{M}(\PC,1) \label{eqn11}\\
\PC\sim \mathcal{D}(\alpha)\label{eqn2}\\
\PC\vert X\sim \mathcal{D}(\alpha+\widehat{\PC}^{ML})\label{eqn3}
\end{eqnarray}
where $\mathcal{D}(\alpha)$ denotes the Dirichlet distribution with parameter $\alpha$, a vector with the same dimension as $\PC$ and $\widehat{\PC}^{ML}$ is the maximum likelihood for $\PC$, corresponding to the observed proportions of each combination in the data set. 
 A classical choice for $\alpha$ is $\alpha=(1/2,\ldots,1/2)$ corresponding to the non-informative Jeffreys prior \cite{Box92}. Combining the prior distribution and the observed entries, a posterior distribution for the model's parameters is obtained (Equation (\ref{eqn3})).

Because missing values occur in the data set, the posterior distribution is not tractable, therefore, drawing a set of model's parameters in it is not straightforward. Thus, a data-augmentation algorithm \cite{Tanner87} is used. In the first step of the algorithm, missing values are imputed by random values. Then, because the data set is now completed, a draw of $\PC$ in the posterior distribution \eqref{eqn3} can easily be obtained. Next, missing values are imputed from the predictive distribution  \eqref{eqn11} using  the previously drawn  parameter and the observed values. These steps of imputation and draw from the posterior distribution are repeated until convergence. At the end, one set of parameters $\tilde{\PC}_{\nbtab}$, drawn from the observed posterior distribution, is obtained.
Repeating the procedure $\Nbtab$ times in parallel, $\Nbtab$ sets of parameters are obtained from which multiple imputation can be done. In this way, the uncertainty on the parameters of the imputation model is reflected, insuring a proper imputation.

The loglinear model is considered as the gold standard for MI of categorical data \cite{Vermunt08}. Indeed, this imputation model reflects all kind of relationships between variables, which enables applying any analysis model. However, this method is dedicated to data sets with a small number of categories because it requires a number of independent parameters equal to the number of combinations of categories minus 1. For example, it corresponds to 9 765 624 independent parameters for a data set with $\Nbvar=10$ variables with $q_k=5$ categories for each of them. This involves two issues: the storage of $\PC$ and overfitting. To overcome these issues, the model can be simplified by adding constraints on $\PC$. The principle is to write $log(\PC)$ as a linear combination of a restricted set of parameters $\lambda=\left[\lambda_0,\lambda_{x_1},\ldots,\lambda_{x_{\Nbvar}},\ldots,\lambda_{x_1 x_2},\ldots,\lambda_{x_1 x_{\Nbvar}},\ldots,\lambda_{x_{\Nbvar-1} x_{\Nbvar}} \right]$, where each element is indexed by a category or a couple of categories. More precisely, the constraints on $\PC$ are given by the following equation:
\begin{eqnarray}
log(\PC_{x_1\ldots x_{\Nbvar}})&=&\lambda_0 +\sum_{\nbvar}{\lambda_{x_{\nbvar}}}+
\sum_{\substack{\left(\nbvar,\nbvar '\right)\\
\nbvar \neq \nbvar '}
}{\lambda_{x_{\nbvar}x_{\nbvar '}}}
\label{eqn4}\text{ for all }(X_1=x_1,\ldots,X_{\Nbvar}=x_{\Nbvar})
\end{eqnarray}
where the second sum is the sum over all the couples of categories possible from the set of categories $(x_1,\ldots,x_{\Nbvar})$.
Thus, the imputation model reflects only the simple (two-way) associations between variables, which is generally sufficient. 
Equation (\ref{eqn4}) leads to 760 independent parameters for the previous example. However, although it requires a smaller number of parameters, the imputation under the loglinear model still remains difficult in this case, because the data-augmentation algorithm used \cite[p.320]{Schafer97} is based on a modification of $\PC$ at each iteration and not of $\lambda$. Thus the storage issue remains.


\subsection{Multiple imputation using a latent class model}
To overcome the limitation of MI using the loglinear model, another JM method based on the latent class model can be used. The latent class model \cite[p.535]{Agresti02} is a mixture model based on the assumption that each individual belongs to a latent class from which all variables can be considered as independent. More precisely, let $\Varnc$ denote the latent categorical variable whose values are in $\lbrace 1,\ldots,\Nc\rbrace$. Let $\PC_{\Varnc}=\left(\PC_{\nc}\right)_{1\leq \nc \leq \Nc}$ denote the proportion of the mixture and $\PC_X=\left(\PC_{x}^{(\nc)}\right)_{1\leq \nc \leq \Nc}$ the parameters of the $\Nc$ components of the mixture. Thus, let $\PC=\left(\PC_{\Varnc}, \PC_{X}\right)$ denote the parameters of the mixture, the joint distribution of the data is written as follows:
\begin{eqnarray}
\mathbb{P}\left(X=(x_1,\ldots,x_{\Nbvar}); \PC \right)&=&
\sum_{\nc=1}^{\Nc}{\left(
\mathbb{P}\left(\Varnc=\nc, \PC_{\Varnc}\right)\prod_{\nbvar=1}^{\Nbvar}{\mathbb{P}\left(X_{\nbvar}=x_{\nbvar}\vert\Varnc=\nc;\PC_{x}^{(\nc)}\right)}\right)
}
\label{eqnCL}
\end{eqnarray}
Assuming a multinomial distribution for $Z$ and $X\vert Z$, Equation (\ref{eqnCL}), can be rewritten as follows:
\begin{eqnarray}
\mathbb{P}\left(X=(x_1,\ldots,x_{\Nbvar}); \PC \right)&=&
\sum_{\nc=1}^{\Nc}{\left(
\PC_{\nc}\prod_{\nbvar=1}^{\Nbvar}{\PC_{x_{\nbvar}}^{(\nc)}}
\right)}
\end{eqnarray}

The latent class model requires $\Nc \times \left(\Nbcol - \Nbvar\right) + \left(\Nbvar - 1 \right)$ independent parameters,\textit{ i.e.} a number that linearly increases with the number of categories.

\cite{Vidotto14} reviews in detail different multiple imputation methods using a latent class model. These methods can be distinguished by the way used to reflect the uncertainty on the parameters of the imputation model and by the way that the number of components of the mixture is chosen: automatically or \textit{a priori}. The quality of the imputation is quite similar from one method to another, the main differences remain in computation time.  One of the latest contributions in this family of methods uses a non-parametric extension of the model namely the Dirichlet process mixture of products of multinomial distributions model (DPMPM) \cite{Dunson09, Si2013}. This method uses a fully Bayesian approach in which the number of classes is defined automatically and is not too computationally intensive. DPMPM assumes a prior distribution on $\PC_{\Varnc}=\left(\PC_{\nc}\right)_{1\leq \nc \leq \Nc}$ without fixing the number of classes which is supposed to be infinite. More precisely, the prior distribution for $\PC_{\Varnc}$ is defined as follows:
\begin{eqnarray}
\PC_{\nc}&=&\zeta_{\nc}\prod_{g<\nc}(1-\zeta_g) \text{	for }\nc \text{ in }1,\ldots,\infty \\
\zeta_{\nc}&\sim& \mathcal{B}(1,\alpha)\\
\alpha&\sim&\mathcal{G}(.25,.25)
\end{eqnarray}
where $\mathcal{G}$ refers to the gamma distribution, $\alpha$ is a positive real number, $\mathcal{B}$ refers to the beta distribution;  the prior distribution for $\PC_{X}$ is defined by:
\begin{eqnarray}
\PC_x^{(\nc)}\sim \mathcal{D}\left(1,\ldots,1\right)
\end{eqnarray}
corresponding to a uniform distribution over the simplex defined by the constraint of sum to one.
The posterior distribution of $\PC$ is not analytically tractable, even when no missing value occur. However, the distribution of each parameter is known if the others are given. For this reason, a Gibbs sampler is used to obtain a draw from the posterior distribution. The principle of this is to draw each parameter while fixing the others. From an incomplete data set, missing values require to be preliminarily imputed. More precisely, a draw from the posterior distribution is obtained as follows: first, the parameters and missing values are initialized; then, given the current parameters, particularly $\PC_{\Varnc}$ and $\PC_{X}$, each individual is randomly affected to one class according to its categories; next, each parameter ($\PC_{\Varnc}$, $\PC_X$, $\alpha$) is drawn conditionally to the others; finally, missing values are imputed according to the mixture model. These steps are then repeated until convergence (for more details, see \cite{Si2013}).

Despite the infinite number of classes, the prior on $\PC_{\nc}$ typically implies  that the posterior distribution for $\PC_{\nc}$ is non negligible for a finite number of classes only. Moreover, for computational reasons, the number of classes has to be bounded. Thus, \cite{Si2013} recommends to fix the maximum number of latent classes to twenty. Consequently, the simulated values of $\PC$ are some realisations of an approximated posterior distribution only.

Multiple imputation using the latent class model has the advantages and drawbacks of this model: because the latent class model approximates quite well any kind of relationships between variables, MI using this model enables the use of complex analysis models such as logistic regression with some interaction terms and provides good estimates of the parameters of the analysis model. However, the imputation model implies that given a class, each individual is imputed in the same way, whatever the categories taken. If the class is very homogeneous, all the individuals have the same observed values, and this behaviour makes sense. However, when the number of missing values is high and when the number of variables is high, it is not straightforward to obtain homogeneous classes. It can explain why \cite{Vidotto14} observed that the multiple imputation using the latent class model can lead to biased estimates for the analysis model in such cases.

 

\subsection{Multiple imputation using a multivariate normal distribution\label{norm}}
Another popular strategy to perform MI for categorical data is to adapt the methods developed for continuous data. Because multiple imputation using the normal multivariate distribution is a robust method for imputing continuous non-normal data \cite{Schafer97}, imputation using the multivariate normal model is an attractive method for this. The principle consists in recoding the categorical variables as dummy variables and applying the multiple imputation under the normal multivariate distribution on the recoded data. The imputed dummy variables are seen as a set of latent continuous variables from which categories can be independently derived. More precisely, let $\bfZ_{\Nbind \times \Nbcol}$ denote the disjunctive table coding for $\bfX_{\Nbind\times \Nbvar}$, \textit{i.e.}, the set of dummy variables corresponding to the incomplete matrix. Note that one missing value on $\bfx_{\nbvar}$ implies $q_{\nbvar}$ missing values for $\bfz_{\nbvar}$. The following procedure implemented in \cite{ameliapackage,Honaker11} enables the multiple imputation of a categorical data set using the normal distribution:
\begin{itemize}
\item perform a non-parametric bootstrap on $\bfZ$: sample the rows of $\bfZ$ with replacement $\Nbtab$ times. $\Nbtab$ incomplete disjunctive tables $\left(\bfZ^{boot}_{\nbtab}\right)_{1\leq\nbtab\leq\Nbtab}$ are obtained;
\item estimate the parameters of the normal distribution on each bootstrap replicate: calculate the ML estimators of $\left(\mu_{\nbtab},\Sigma_{\nbtab}\right)$, the mean and the variance of the normal distribution for the $\nbtab^{th}$ bootstrap incomplete replicate, using an EM algorithm. Note that the set of $\Nbtab$ parameters reflects  the uncertainty required for a proper multiple imputation method;
\item create $\Nbtab$ imputed disjunctive tables: impute $\bfZ$ from the normal distribution using  $\left(\mu_{\nbtab},\Sigma_{\nbtab}\right)_{1\leq\nbtab\leq\Nbtab}$ and the observed values of $\bfZ$. $\Nbtab$ imputed disjunctive tables $\left(\bfZ_{\nbtab}\right)_{1\leq\nbtab\leq\Nbtab}$ are obtained. In $\bfZ_{\nbtab}$, the observed values are still zeros and ones, whereas the missing values have been replaced by real numbers;
\item create $\Nbtab$ imputed categorical data sets: from the latent continuous variables contained in $\left(\bfZ_{\nbtab}\right)_{1\leq\nbtab\leq\Nbtab}$, derive categories for each incomplete individual.
\end{itemize}
Several ways have been proposed to get the imputed categories from the imputed continuous values. For example \cite{Allison02} recommends to attribute the category corresponding to the highest imputed value, while \cite{Bernaards07,Demirtas09,Yucel08} propose some rounding strategies. However, \textit{``A single best rounding rule for categorical data has yet to be identified."} \cite[p. 107]{VB12}. A common one proposed by \cite{Bernaards07} is called \textit{Coin flipping}. Coin flipping consists in considering the set of imputed values of the $q_{\nbvar}$ dummy variables $\bfz_{\nbvar}$ as an expectation given the observed values $ \PC_{\nbvar}=\mathbb{E}\left[(z_{1},\ldots,z_{q_{\nbvar}})|Z_{obs}; \hat{\mu},\hat{\Sigma}\right]$. Thus, randomly drawing one category according to a multinomial distribution $\mathcal{M}\left(\PC_{\nbvar},1\right)$, suitably modified so that $\PC_{\nbvar}$ remains between 0 and 1, imputes plausible values. The values lower than 0 are replaced by 0 and the imputed values higher than 1 are replaced by 1. In this case, the imputed values are scaled to respect the constraint of sum to one.

Because imputation under the normal multivariate distribution is based on the estimate of a covariance matrix, the imputation under the normal distribution can detect only two-way associations between categorical variables. In addition, this method assumes independence between categories conditionally to the latent continuous variables. This implies that if two variables are linked, and if an individual has missing values on these ones, then the categories derived from the imputed disjunctive table will be drawn independently.
Consequently, the two-way associations can not be perfectly reflected in the imputed data set. Note that, contrary to the MI using the latent class, the parameter of the multinomial distribution $\PC_{\nbvar}$ is specific to each individual, because the imputation of the disjunctive table is performed given the observed values. This behaviour makes sense if the variables on which missing values occur are linked with the others.
The main drawback of the MI using the normal distribution is the number of independent parameters estimated. This number is equal to $\frac{\left(\Nbcol-\Nbvar\right)\times\left(\Nbcol-\Nbvar+1\right)}{2}+\left(\Nbcol-\Nbvar\right)$, representing 860 parameters for a data set with 10 variables with 5 categories. It increases rapidly when the total number of categories ($\Nbcol$) increases, leading quickly to overfitting. Moreover, the covariance matrix is not invertible when the number of individuals is lower than $\left(\Nbcol-\Nbvar\right)$. To overcome these issues, it is possible to add a ridge term on its diagonal to improve the conditioning of the regression problem.

\subsection{Fully conditional specification\label{fcs}}
Categorical data can be imputed using a FCS approach instead of a JM approach: for each variable with missing values, an imputation model is defined, (\textit{i.e.} a conditional distribution), and each incomplete variable is sequentially imputed according to this, while reflecting the uncertainty on the model's parameters. Typically, the models used for each incomplete variable are some multinomial logistic regressions and the variability of the models parameter is reflected using a Bayesian point of view. More precisely, we denote by $\PC_{{\nbvar}}=\left(\PC_{k_{\nc}}\right)_{1\leq\nc\leq q_k}$ the set of parameters for the multinomial distribution of the variable to impute $X_{\nbvar}$ (the set of the other variables is denoted $X_{-\nbvar}$). We also denote by
$\boldsymbol{\beta}_{\nbvar}=\left(\boldsymbol{\beta_{\nbvar 1}},\ldots,\boldsymbol{\beta_{\nbvar \Nc}}\right)$ the set of regression parameters that defines $\PC_{{\nbvar}}$, such as $\boldsymbol{\beta}_{\nbvar \nc}$ is the regression parameter vector associated with the category $\nc$ of the response variable $X_{\nbvar}$ and $\bfZ_{\nbvar}$ is the design matrix associated. Note that identifiability constraints are required on $\boldsymbol{\beta}_{\nbvar}$, that is why $\boldsymbol{\beta}_{\nbvar \Nc}$ is fixed to the null vector. 
Thus, the imputation is built on the following assumptions:
\begin{eqnarray}
X_{\nbvar}\vert\PC_{\nbvar}  \sim \mathcal{M}(\PC_{\nbvar},1)\\
\PC_{\nbvar\nc}=\mathbb{P}(X_k=\nc\vert X_{- \nbvar},\beta)=\frac{exp(\bfZ_{\nbvar} \boldsymbol{\beta}_{\nbvar\nc})}{1+ \sum_{\nc=1}^{\Nc -1}exp(\bfZ_{\nbvar}\boldsymbol{\beta}_{\nbvar\nc})}\\
\beta\vert X \sim\mathcal{N}(\hat\beta, \hat V)
\end{eqnarray}
where $\hat\beta, \hat V$  are the estimators of $\beta$ and of its associated variance.
For simplicity, suppose that the data set contains 2 binary variables $\bfx_1$ and $\bfx_2$, with $\bfx_2$ as incomplete and $\bfx_1$ as complete. To impute $\bfx_2$ given $\bfx_1$ the first step is to estimate $\beta$ and its associated variance using complete cases by iteratively reweighted least squares.
Then, a new parameter $\tilde{\beta}_{\nbvar}$ is drawn from a normal distribution centred in the previous estimate with the covariance matrix previously obtained. Lastly, the fitted probability $\PC_{\nbvar}$ are obtained from the logistic regression model with parameter $\tilde{\beta}_{\nbvar}$ and $\bfx_2$ is imputed according to a multinomial distribution with parameters $\PC_{\nbvar}$ \cite[p.76]{VB12}. Note that $\beta$ is drawn in an approximated posterior distribution. Indeed, as explained by \cite[p.169-170]{Rubin87}, the posterior distribution has not a neat form for reasonable prior distributions. However, on a large sample, assuming a weak prior on $\beta$, the posterior distribution can be approximated by a normal distribution. Thus, draw $\beta$ in a normal distribution with $\hat\beta$ and $\hat V$ as parameters makes sense.

In the general case, where the data set contains $\Nbvar$ variables with missing values, each variable is imputed according to a multinomial logistic regression given all the others. More precisely, the incomplete data set is firstly randomly imputed. Then, the missing values of the variable $\bfx_{\nbvar}$ are imputed as explained previously: a value of $\beta_{\nbvar}$ is drawn from the approximated posterior distribution and an imputation according to $\mathbb{P}\left(X_{\nbvar}\vert X_{-\nbvar};{\PC}_{{\nbvar}} \right)$ is performed. The next incomplete variable is imputed in the same way given the other variables, and particularly from the new imputed values of $\bfx_{\nbvar}$. We proceed in this way for all variables and repeat it until convergence, this provides one imputed data set. The procedure is performed $\Nbtab$ times in parallel to provide $\Nbtab$ imputed data sets.

Implicitly, the choices of the conditional distributions $\mathbb{P}\left(X_{\nbvar}\vert X_{-\nbvar};\PC_{{\nbvar}} \right)$ determine a joint distribution $\mathbb{P}\left(X_{\nbvar};\PC\right)$, in so far as a joint distribution is compatible with these choices \cite{Besag74}. The convergence to the joint distribution 
is often obtained for a low number of iterations (5 can be sufficient), but \cite[p.113]{VB12} underlines that this number can be higher in some cases. In addition, FCS is more computationally intensive than JM \cite{VB12,Vermunt08}. This is not a practical issue when the data set is small, but it becomes so on a data set of high dimensions. In particular, checking the convergence becomes very difficult.

The imputation using logistic regressions on each variable performs quite well, that is why this method is often used as a benchmark to perform comparative studies \cite{vanderPalm14,Doove14, Shah14,Si2013}. However, the lack of multinomial regression can affect the multiple imputation procedure using this model. Indeed, when separability problems occur \cite{Albert84}
, or when the number of individuals is smaller than the number of categories \cite[p.195]{Agresti02}, it is not possible to get the estimates of the parameters. In addition, the number of parameters is very large when the number of categories per variable is high, implying overfitting when the number of individuals is small. When the number of categories becomes too large, \cite{VB12,vanBuuren11} advise to use a method dedicated to continuous data: the predictive mean matching (PMM). PMM treats each variable as continuous variables, predicts them using linear regression, and draws one individual among those the nearest to the predicted value. However, PMM often yields to biased estimates \cite{vanderPalm14}.

Typically, the default models selected for each logistic regression are main effects models. Thus, the imputation model captures the two-way associations between variables well, which is generally sufficient for the analysis model. However, models taking into account interactions can be used but the choice of these models requires a certain effort by the user. To overcome this effort, in particular when the variables are numerous, conditional imputations using random forests instead of logistic regression have been proposed \cite{Doove14, Shah14}. According to \cite{Doove14}, an imputation of one variable $X_{\nbvar}$ given the others is obtained as follows:
\begin{itemize}
\item draw 10 bootstrap samples from the individuals without missing value on $X_{\nbvar}$;
\item fit one tree on each sample: for a given bootstrap sample, draw randomly a subset of $\sqrt{\Nbvar-1}$ variables among the $\Nbvar-1$ explanatory variables. Build one tree from this bootstrap sample and this subset of explanatory variables. A random forest of 10 trees is obtained. Note that the uncertainty due to missing values is reflected by the use of one random forest instead of a unique tree;
\item impute missing values on $X_{\nbvar}$ according to the forest: for an individual $\nbind$ with a missing value on $X_{\nbvar}$, gather all the donors from the 10 predictive leaves from each tree and draw randomly one donor from it.
\end{itemize}
Then, the procedure is performed for each incomplete variable and repeated until convergence. Using random forests as conditional models allows capturing complex relationships between variables. In addition, the method is very robust to the number of trees used, as well as to the number of explanatory variables retained. Thus, the default choices for these parameters (10 trees, $\sqrt{\Nbvar-1}$ explanatory variables) are very suitable in most of the cases. However, the method is more computationally intensive than the one based on logistic regressions.

\section{Multiple Imputation using multiple correspondence analysis\label{sec3}}
This section deals with a novel MI method for categorical data based on multiple correspondence analysis (MCA) \cite{Green06,Leb84}, \textit{i.e.} the principal components method dedicated for categorical data. Like the imputation using the normal distribution, it is a JM method based on the imputation of the disjunctive table. We first introduce MCA as a specific singular value decomposition on specific matrices. Then, we present how to perform this SVD with missing values and how it is used to perform single imputation. We explain how to introduce uncertainty to obtain a proper MI method. Finally, the properties of the method are discussed and the differences with MI using the normal distribution highlighted.
\subsection{MCA for complete data\label{secMCA}}
MCA is a principal components method to describe, summarise and visualise multidimensional matrices with categorical data. This powerful method allows us to understand the two-way associations between variables as well as the similarities between individuals. Like any principal components method, MCA is a method of dimensionality reduction consisting in searching for a subspace of dimension $\Nbdim$ providing the best representation of the data in the sense that it 
maximises the variability of the projected points (\textit{i.e.} the individuals or the variables according to the space considered). The subspace can be obtained by performing a specific singular value decomposition (SVD) on the disjunctive table.

More precisely, let $\bfZ_{\Nbind \times \Nbcol}$ denote the disjunctive table corresponding to $\bfX_{\Nbind\times \Nbvar}$. We define a metric between individuals through the diagonal matrix $\frac{1}{\Nbvar}\bfD_{\Sigma}^{-1}$ where

\noindent $\bfD_{\Sigma}=\bf{diag} \left(p^{x_{1}}_{1},\ldots,p^{x_{1}}_{q_1},\ldots,p^{x_{\Nbvar}}_{1},\ldots,p^{x_{\Nbvar}}_{q_{\Nbvar}}\right)$ is a diagonal matrix with dimensions $\Nbcol\times \Nbcol$, $p^{\bfx_{\nbvar}}_{\ell}$ is the proportion of observations taking the category $\ell$ on the variable $\bfx_{\nbvar}$. In this way, two individuals taking different categories for the same variable are more distant from the others when one of them takes a rare category than when both of them take frequent categories. We also define a uniform weighting for the individuals through the diagonal matrix $\frac{1}{\Nbind}\bbone_{\Nbind}$ with $\bbone_{\Nbind}$ the identity matrix of dimensions $\Nbind$. By duality, the matrices $\frac{1}{\Nbvar}\bfD_{\Sigma}^{-1}$ and $\frac{1}{\Nbind}\bbone_{\Nbind}$ define also a weighting and a metric for the space of the categories respectively.
MCA consists in searching a matrix $\widehat{\bfZ}$ with a lower rank $\Nbdim$ as close as possible to the disjunctive table $\bfZ$ in the sense defined by these metrics. 
Let $\bfM_{\Nbind \times \Nbcol}$ denote the matrix where each row is equal to the vector of the means of each column of $\bfZ$. 
MCA consists in performing the SVD of the matrix triplet $\left(\bfZ-\bfM,
\frac{1}{\Nbvar}\bfD_{\Sigma}^{-1},
\frac{1}{\Nbind}\bbone_{\Nbind}
\right)$ \cite{Green84} which is equivalent to writing $\left(\bfZ-\bfM\right)$ as
\begin{eqnarray}
\bfZ-\bfM=\bfU\bf{\Lambda}^{1/2}\bfV^{\top}
\end{eqnarray}
where the columns of $\bfU_{\Nbind\times \Nbcol}$ are the left singular vectors satisfying the relationship

\noindent $\bfU^{\top}\bf{diag}(1/{\Nbind},\ldots,1/{\Nbind})\bfU=\bbone_{\Nbcol}$; columns of $\bfV_{\Nbcol\times \Nbcol}$ are the right singular vectors satisfying the relationship $\bfV^{\top}\frac{1}{\Nbvar}\bfD_{\Sigma}^{-1}\bfV=\bbone_{\Nbcol}$ and
 $\bf{\Lambda}^{1/2}_{\Nbcol\times\Nbcol}=diag\left(\lambda_1^{1/2},\ldots,\lambda_{\Nbcol}^{1/2}\right)$ is the diagonal matrix of the singular values.
 
The $\Nbdim$ first principal components are given by $\widehat{\bfU}_{\Nbind\times \Nbdim}\widehat{\bf{\Lambda}}^{1/2}_{\Nbdim\times \Nbdim}$, the product between the first columns of $\bfU$ and the diagonal matrix $\bf{\Lambda}^{1/2}$ restricted to its $\Nbdim $ first elements. In the same way, the $\Nbdim$ first loadings are given by $\widehat{\bfV}_{\Nbcol\times \Nbdim}$. $\widehat{\bfZ}$  defined by:
\begin{eqnarray}
\widehat{\bfZ}=\widehat{\bfU}\widehat{\bf{\Lambda}}\widehat{\bfV}^{\top}+\bfM
\label{reconst}
\end{eqnarray}
is the best approximation of $\bfZ$, in the sense of the metrics, with the constraint of rank $\Nbdim$ (Eckart-Young theorem \cite{Eckhart36}). Equation (\ref{reconst}) is called \textit{reconstruction formula}.

Note that, contrary to $\bfZ$, $\bf\widehat{Z}$ is a fuzzy disjunctive table in the sense that its cells are real numbers and not only zeros and ones as in a classic disjunctive table. However, the sum per variable is still equal to one \cite{Tenenhaus85}. Most of the values are contained in the interval $\left[0,1\right]$ or close to it because $\bf\widehat{Z}$ is as close as possible to $\bfZ$ which contains only zeros and ones, but values out of this interval can occur.

Performing MCA requires $\Nbcol-\Nbvar$ parameters corresponding to the terms useful for the centering and the weighting of the categories, $\Nbind\Nbdim-\Nbdim-\frac{\Nbdim(\Nbdim+1)}{2}$ for the centered and orthonormal left singular vectors and  $(\Nbcol-\Nbvar)\Nbdim-\Nbdim-\frac{\Nbdim(\Nbdim+1)}{2}$ for the orthonormal right singular vectors, for a total of
 $\Nbcol-\Nbvar+\Nbdim\left(\Nbind-1+\left(\Nbcol-\Nbvar\right)-\Nbdim\right)$ independent parameters. This number of parameters increases linearly with the number of values in the data set.
 
\subsection{Single imputation using MCA}
\cite{Josse12} proposed an iterative algorithm called ``iterative MCA'' to perform single imputation using MCA. The main steps of the algorithm are as follows: 
\begin{enumerate}
\item initialization $\ell=0$: recode $\bfX$ as disjunctive table $\bfZ$, substitute missing values by initial values (the proportions) and calculate $\bfM^0$ and $\bfD_{\Sigma}^0$ on this completed data set.
\item step $\ell$:
\begin{itemize}
\item[(a)] perform the MCA, in other words the SVD of $\left(\bfZ^{\ell-1}-\bfM^{\ell-1},\frac{1}{\Nbvar}\left(\bfD_{\Sigma}^{{\ell-1}}\right)^{-1},\frac{1}{\Nbind}\bbone_{\Nbind}\right)$ 
to obtain $\hat \bfU^{\ell}$, $\hat \bfV^{\ell}$ and $\left(\hat{\bf\Lambda}^{\ell}\right)^{1/2}$;
\item[(b)] keep the $S$ first dimensions and use the reconstruction formula (\ref{reconst}) to compute the fitted matrix: 
$$\hat{\bfZ}^{\ell}_{I\times J} = \left(\hat{\bfU}_{I\times S}^{\ell} \left(\hat{\bf\Lambda}_{S\times S}^{\ell}\right)^{1/2}\left(\hat{\bfV}_{J\times S}^{\ell}\right)^{\top}\right)+\bfM^{\ell -1}_{I\times J}$$
and the new imputed data set becomes $\bfZ^{\ell}= \bfW *\bfZ + (\bbone-\bfW)* \hat \bfZ^{\ell}$ with $*$ being the Hadamard product, ${\bbone}_{I\times J}$ being a matrix with only ones and $\bfW$ a weighting matrix where $w_{\nbind\nbcol}=0$ if $z_{\nbind\nbcol}$ is missing and $w_{\nbind\nbcol}=1$ otherwise. The observed values are the same but the missing ones are replaced by the fitted values;
\item[(c)] from the new completed matrix $\bfZ^{\ell}$, $\bfD_{\Sigma}^{\ell}$ and $\bfM^{\ell}$ are updated.
\end{itemize}
\item steps (2.a), (2.b) and (2.c) are repeated until the change in the imputed matrix falls below a predefined threshold $\sum_{ij}(\hat z_{\nbind\nbcol}^{\ell-1}-\hat z_{\nbind\nbcol}^{\ell})^2\leq \varepsilon$, with $\varepsilon$ equals to $10^{-6}$ for example.
\end{enumerate}
The iterative MCA algorithm consists in recoding the incomplete data set as an incomplete disjunctive table, randomly imputing the missing values, estimating the principal components and loadings from the completed matrix and then, using these estimates to impute missing values according to the reconstruction formula (\ref{reconst}). The steps of estimation and imputation are repeated until convergence, leading to an imputation of the disjunctive table, as well as to an estimate of the MCA parameters.

The algorithm can suffer from overfitting issues, when missing values are numerous, when the relationships between variables are weak, or when the number of observations is low. To overcome these issues, a regularized version of it has been proposed \cite{Josse12}. The rationale is to remove the noise in order to avoid instabilities in the prediction by replacing the singular values $\left(
\sqrt{\hat{\bf\lambda}_{\nbdim}^{\ell}}\right)_{1\leq\nbdim\leq\Nbdim}$ of step (2.b) by \textit{shrunk} singular values $\left(\frac{\hat{\bf\lambda}_{\nbdim}^{\ell}-
\sum_{\nbdim=\Nbdim+1}^{\Nbcol-\Nbvar}{\frac{\lambda_{\nbdim}}{\Nbcol-\Nbvar-\Nbdim}}
}{
\sqrt{\hat{\bf\lambda}_{\nbdim}^{\ell}}
}
\right)_{1\leq\nbdim\leq\Nbdim}$.
In this way, singular values are thresholded with a greater amount of shrinkage for the smallest ones. Thus, the first dimensions of variability take a more significant part in the reconstruction of the data than the others. Assuming that the first dimensions of variability are made of information and noise, whereas the last ones are made of noise only, this behaviour is then satisfactory. Geometrically, the regularization makes the individual closer to the center of gravity. Concerning the cells of $\widehat{\bfZ}$, the regularization makes the values closer to the mean proportions  and consequently, these values are more often in the interval $\left[0,1\right]$.\\

The regularized iterative MCA algorithm enables us to impute an incomplete disjunctive table but not an initial incomplete data set. A strategy to go from the imputed disjunctive table to an imputed categorical data set is required. We also suggest the use of the coin flipping approach. Let us note that for each set of dummy variables coding for one categorical variable, the sum per row is equal to one, even if it contains imputed values. Moreover, most of the imputed cells are in the interval $\left[0,1\right]$ or are close to it. Consequently, modifications of these cells are not often required.
\subsection{MI using MCA}

To perform MI using MCA, we need to reflect the uncertainty concerning the principal components and loadings. To do so, we use a non-parametric bootstrap approach based on the specificities of MCA. Indeed, as seen in Section \ref{secMCA}, MCA enables us to assign a weight to each individual. This possibility to include a weight for the individual is very useful when the same lines of the data set occur several times. Instead of storing each replicate, a weight proportional to the number of occurrences of each line can be used, allowing the storage only of the lines that are different.
Thus, a non-parametric bootstrap, such as the one used for the MI using the normal distribution, can easily be performed simply by modifying the weight of the individuals: if an individual does not belong to the bootstrap replicate, then its weight is null,  otherwise, its weight is proportional to the number of times the observation occurs in the replicate.
Note that individuals with a weight equal to zero are classically called \textit{supplementary individuals} in the MCA framework \cite{Green84}.

Thus, we define a MI method called multiple imputation using multiple correspondence analysis (MIMCA). First, the algorithm consists in drawing $\Nbtab$ sets of weights for the individuals. Then, $\Nbtab$ single imputations are performed: at first, the regularized iterative MCA algorithm is used to impute the incomplete disjunctive table using the previous weighting for the individuals; Next, coin flipping is used to obtain categorical data and mimic the distribution of the categorical data. At the end, $\Nbtab$ imputed data sets are obtained and any statistical method can be applied on each one. In detail, the MIMCA algorithm is written as follows:

\begin{enumerate}
	\item Reflect the variability on the set of parameters of the imputation model: draw $\Nbind$ values with replacement in $\lbrace 1,..,\Nbind \rbrace$ and define a weight $r_{\nbind}$ for each individual proportional to the number of times the individual $\nbind$ is drawn. 
	\item Impute the disjunctive table according to the previous weighting:
	\begin{enumerate}
		\item initialization $\ell=0$: recode $\bfX$ as a disjunctive table $\bfZ$, substitute missing values by initial values (the proportions) and calculate $\bfM^0$ and $\bfD_{\Sigma}^0$ on this completed data set.
		\item step $\ell$:
		\begin{enumerate}
			\item perform the SVD of $\left(\bfZ^{\ell-1}-\bfM^{\ell-1},\frac{1}{\Nbvar}\left(\bfD_{\Sigma}^{{\ell-1}}\right)^{-1},{\bf{diag}}\left(r_1,\ldots,r_{\Nbind}\right)\right)$ 
to obtain $\hat \bfU^{\ell}$, $\hat \bfV^{\ell}$ and $\left(\hat{\bf\Lambda}^{\ell}\right)^{1/2}$;
			\item keep the $S$  first dimensions and compute the fitted matrix: 
$$\hat{\bfZ}^{\ell}= \left(\hat{\bfU}^{\ell} \left(\hat{\bf\Lambda}^{\ell}_{shrunk}\right)^{1/2}\left(\hat{\bfV}^{\ell}\right)^{\top}\right)+\bfM^{\ell -1}$$
where $\left(\hat{\bf\Lambda}^{\ell}_{shrunk}\right)^{1/2}$ is the diagonal matrix containing the shrunk singular values and derive the new imputed data set $\bfZ^{\ell}= \bfW *\bfZ + (\bf1-\bfW)* \hat \bfZ^{\ell}$
			\item from the new completed matrix $\bfZ^{\ell}$, $\bfD_{\Sigma}^{\ell}$ and $\bfM^{\ell}$ are updated.
		\end{enumerate}
		\item step (2.b) is repeated until convergence.
	\end{enumerate}
	\item Mimic the distribution of the categorical data set using coin flipping on $\bfZ^{\ell}$ :
	\begin{enumerate}
	\item if necessary, modify suitably the values of $\bfZ^{\ell}$: negative values are replaced by zero, and values higher than one are replaced by one. Then, for each set of dummy variables coding for one categorical variable, scale in order to verify the constraint that the sum is equal to one.
	\item for imputed cells coding for one missing value, draw one category according to a multinomial distribution.
	\end{enumerate}
\item Create $\Nbtab$ imputed data sets: for $\nbtab$ from 1 to $\Nbtab$ alternate steps 1, 2 and 3.
\end{enumerate}
\subsection{Properties of the imputation method\label{secprop}}
MI using MCA is part of the family of joint modelling MI methods, which means that it avoids the runtime issues of conditional modelling. Most of the properties of the MIMCA method are directly linked to MCA properties. MCA provides an efficient summary of the two-way associations between variables, as well as the similarities between individuals. The imputation benefits from these properties and provides an imputation model sufficiently complex to apply then an analysis model focusing on two-way associations between variables, such as a main effects logistic regression model. In addition, like the MI using the normal distribution, MIMCA uses draws from a multinomial distribution with parameter $\PC$ (obtained by the disjunctive table) specific to each individual and depending on the observed values of the other variables. Lastly, because of the relatively small number of parameters required to perform MCA, the imputation method works well even if the number of individuals is small. These properties have been highlighted in previous works on imputation using principal components methods \cite{Audigier14,Audigier15}.


Since these two methods, MIMCA and the multiple imputation with the normal distribution, provide several imputations of the disjunctive table, and then use the same strategy to go from the disjunctive table to the categorical data set, they seem very close. However, they differ on many other points.

The first one is that the imputation of the disjunctive table by MCA is a deterministic imputation, replacing a missing value by the most plausible value given by the estimate of the principal components and the estimate of the loadings. Then, coin flipping is used to mimic the distribution of the categorical data. On the contrary, the multiple imputation based on the normal distribution uses stochastic regressions to impute the disjunctive table, that is to say, a Gaussian noise is added to the conditional expectation given by the observed values. Then, coin flipping is used, adding uncertainty a second time.


The second difference between the two methods is the covariance of the imputed values. Indeed, the matrix $\widehat{\bfZ}^{\ell}$ contains the reconstructed data by the iterative MCA algorithm and the product $\widehat{\bfZ}^{\ell^{\top}}\widehat{\bfZ}^{\ell}$ provides the covariance matrix of this data. The rank of it is $\Nbdim$. On the contrary, the rank of the covariance matrix used to perform imputation using the normal distribution is $\Nbcol-\Nbvar$ (because of the constraint of the sum equal to one per variable). Consequently, the relationships between imputed variables are different.

The third difference is the number of estimated parameters. Indeed, although the imputation by the normal distribution requires a extremely large number of parameters when the number of categories increases, the imputation using MCA requires a number of parameters linearly dependent to the number of cells. This property is essential from a practical point of view because it makes it very easy to impute data sets with a small number of individuals.

\section{Simulation study}
As mentioned in the introduction, the aim of MI methods is to obtain an inference on a quantity of interest $\psi$. Here, we focus on the parameters of a logistic regression without interaction, which is a statistical method frequently used for categorical data. At first, we present how to make inference for the parameters from multiple imputed data sets. Then, we explain how we assess the quality of the inference built, that is to say, the quality of the MI methods.
Finally, the MI methods presented in Sections \ref{sec2} and \ref{sec3} are compared through a simulation study based on real data sets. It thus provides more realistic performances from a practical point of view. The code to reproduce all the simulations with the R software \cite{Rsoft}, as well as the data sets used, are available on the webpage of the first author.
\subsection{Inference from imputed data sets}
Each MI method gives $\Nbtab$ imputed data sets as outputs. Then, the parameters of the analysis model (for instance the logistic regression) as well as their associated variance are estimated from each one. We denote $\left(\widehat{\psi}_{\nbtab}\right)_{1\leq \nbtab\leq\Nbtab}$ the set of the $\Nbtab$ estimates of the model's parameters and we denote $\left(\widehat{Var}\left(\widehat{\psi}_{\nbtab}\right)\right)_{1\leq \nbtab\leq\Nbtab}$ the set of the $\Nbtab$ associated variances. These estimates have to be pooled to provide a unique estimate of $\psi$ and of its variance using Rubin's rules \cite{Rubin87}.
 
This methodology is explained for a scalar quantity of interest $\psi$. The extension to a vector is straightforward, proceeding in the same way element by element.
The estimate of $\psi$ is simply given by the mean over the $\Nbtab$ estimates obtained from each imputed data set:
\begin{eqnarray}
\hat \psi&=& \frac{1}{\Nbtab}\sum_{\nbtab=1}^{\Nbtab}\hat{\psi}_\nbtab, \label{eqnpool1}
\end{eqnarray}
while the estimate of the variance of $\hat{\psi}$ is the sum of two terms:
\begin{eqnarray}
\widehat{Var}(\hat\psi)&=&\frac{1}{\Nbtab}
\sum_{\nbtab=1}^{\Nbtab}\widehat{Var}\left(\hat{\psi}_{\nbtab}\right)
+\left(1+\frac{1}{\Nbtab}\right)\frac{1}{\Nbtab-1}\sum_{\nbtab=1}^{\Nbtab}{\left(\hat{\psi}_{\nbtab}-\hat{\psi}\right)^2}. \label{eqnpool2}
\end{eqnarray}

The first term is the within-imputation variance, corresponding to the sampling variance. The second one is the between-imputation variance, corresponding to the variance due to missing values. The factor $(1 + \frac{1}{\Nbtab})$ is due to the fact that $\widehat{\psi}$ is estimated from a finite number of imputed tables.

Then, the 95\% confidence interval is calculated as:
\begin{eqnarray*}\hat{\psi}\pm t_{\nu,.975}\sqrt{\widehat{Var}(\hat\psi)}
\end{eqnarray*}
where $t_{\nu,.975}$ is the $.975$ critical value of the Student's $t-$distribution with $\nu$ degrees of freedom estimated as suggested by \cite{Barnard99}.
%

\subsection{Simulation design from real data sets\label{secsimu}}
The validity of MI methods are often assessed by simulation \cite[p.47]{VB12}. We design a simulation study using real data sets to assess the quality of the MIMCA method. Each data set is considered as a population data and denoted $\bfX_{pop}$. The parameters of the logistic regression model are estimated from this population data and they are considered as the true coefficients $\psi$. Then, a sample $\bfX$ is drawn from the population. This step reflects the sampling variance. The values of the response variable of the logistic model are drawn according to the probabilities defined defined by $\psi$. Then, incomplete data are generated completely at random to reflect the variance due to missing values \cite{Brand03}. The MI methods are applied and the inferences are performed. This procedure is repeated $\Nbsim$ times.

The performances of a MI method are measured according to three criteria: the bias given by $
\frac{1}{\Nbsim}\sum_{\nbsim=1}^{\Nbsim}{\left(\hat{\psi}_{\nbsim}-\psi\right)}$, the median (over the $\Nbsim$ simulations) of the confidence intervals width as well as the coverage. This latter is calculated as the percentage of cases where the true value $\psi$ is within its 95\% confidence interval.

A coverage sufficiently close to the nominal level is required to consider that the inference is correct, but it is not sufficient, the confidence interval width should be as small as possible.

To appreciate the value of the bias and of the width of the confidence interval, it is useful to compare them to those obtained from two other methods. The first one consists in calculating the criteria for the data sets without missing values, which we named the ``Full data'' method. The second one is the listwise deletion. This consists in deleting the individuals with missing values. 
Because the estimates of the parameters of the model are obtained from a subsample, the confidence intervals obtained should be larger than those obtained from multiple imputation.

A single imputation method (named \textit{Sample}) is added as a benchmark to understand better how MI methods benefit from using the relationships between variables to impute the data. This single imputation method consists in drawing each category according to a multinomial distribution $\mathcal{M}\left(\theta,1\right)$, with $\theta$ defined according to the proportion of each category of the current variable. 

\subsection{Results}

The methods described in this paper are performed using the following R packages:  \textit{cat} \cite{catpackage} for MI using the saturated loglinear model, \textit{Amelia} \cite{ameliapackage,Honaker11} for MI using a normal distribution, \textit{mi} \cite{mipackage} for MI using the DPMPM method, \textit{mice} \cite{micepackage,vanBuuren11} for the FCS approach using iterated logistic regressions and random forests. This package will also be used to pool the results from the imputed data sets.
The tuning parameters of each MI method are chosen according to their default values implemented in the R packages. Firstly, the tuning parameter of the MIMCA method, that is to say, the number of components, is chosen to provide accurate inferences. Its choice will be discussed later in Section \ref{nbaxe}.

The MI methods are assessed in terms of the quality of the inference as well as the time consumed from data sets covering many situations. The data sets differ in terms of the number of individuals, the number of variables,
the number of categories per variable, the relationships between variables. 


The evaluation is based on the following categorical data sets. For each data set a categorical response variable is available.
\begin{itemize}
\item \textit{Saheart}: This data set \cite{Saheart} provides clinical attributes of $\Nbind_{pop}=462$ males of the Western Cape in South Africa. These attributes can explain the presence of a coronary heart disease. The data set contains $\Nbvar=10$ variables with a number of categories between 2 and 4.
\item \textit{Galetas}: This data set \cite{galetas} refers to the preferences of $\Nbind_{pop}=1192$ judges regarding 11 cakes in terms of global appreciation and in terms of color aspect. The data set contains $\Nbvar=4$ variables with two that have 11 categories.
\item \textit{Sbp}: The $\Nbind_{pop}=500$ subjects of this data set are described by clinical covariates explaining their blood pressure \cite{Sbp}. The data set contains $\Nbvar=18$ variables that have 2 to 4 categories.
\item \textit{Income}: This data set, from the R package \textit{kernlab} \cite{kernlab}, contains $\Nbind_{pop}=6876$ individuals described by several demographic attributes that could explain the annual income of an household. The data set contains $\Nbvar=14$ variables with a number of categories between 2 and 9.
\item \textit{Titanic}: This data set \cite{Titanic} provides information on $\Nbind_{pop}=2201$ passengers on the ocean liner \textit{Titanic}. The $\Nbvar=4$ variables deal with the economic status, the sex, the age and the survival of the passengers. The first variable has four categories, while the other ones have two categories. The data set is available in the R software.
\item \textit{Credit}: German Credit Data from the UCI Repository of Machine Learning Database \cite{UCI} contains $\Nbind_{pop}=982$ clients described by several attributes which enable the bank to classify themselves as good or bad credit risk. The data set contains $\Nbvar=20$ variables with a number of categories between 2 and 4.
\end{itemize}
The simulation design is performed for $\Nbsim=200$ simulations and 20\% of missing values generated completely at random.
The MI methods are performed with $\Nbtab=5$ imputed data sets which is usually enough \cite{Rubin87}. 
\subsubsection{Assessment of the inferences\label{assess}}
First of all, we can note that some methods cannot be applied on all the data sets. As explained previously, MI using the loglinear model can be applied only on data sets with a small number of categories such as \textit{Titanic} or \textit{Galetas}. MI using the normal distribution encounters inversion issues when the number of individuals is small compared to the number of variables. That is why no results are provided for MI using the normal distribution on the data sets \textit{Credit} and \textit{Sbp}. The others MI methods can be applied on all the data sets.

\begin{figure}[h!]
\begin{center}
\includegraphics[scale=.3]{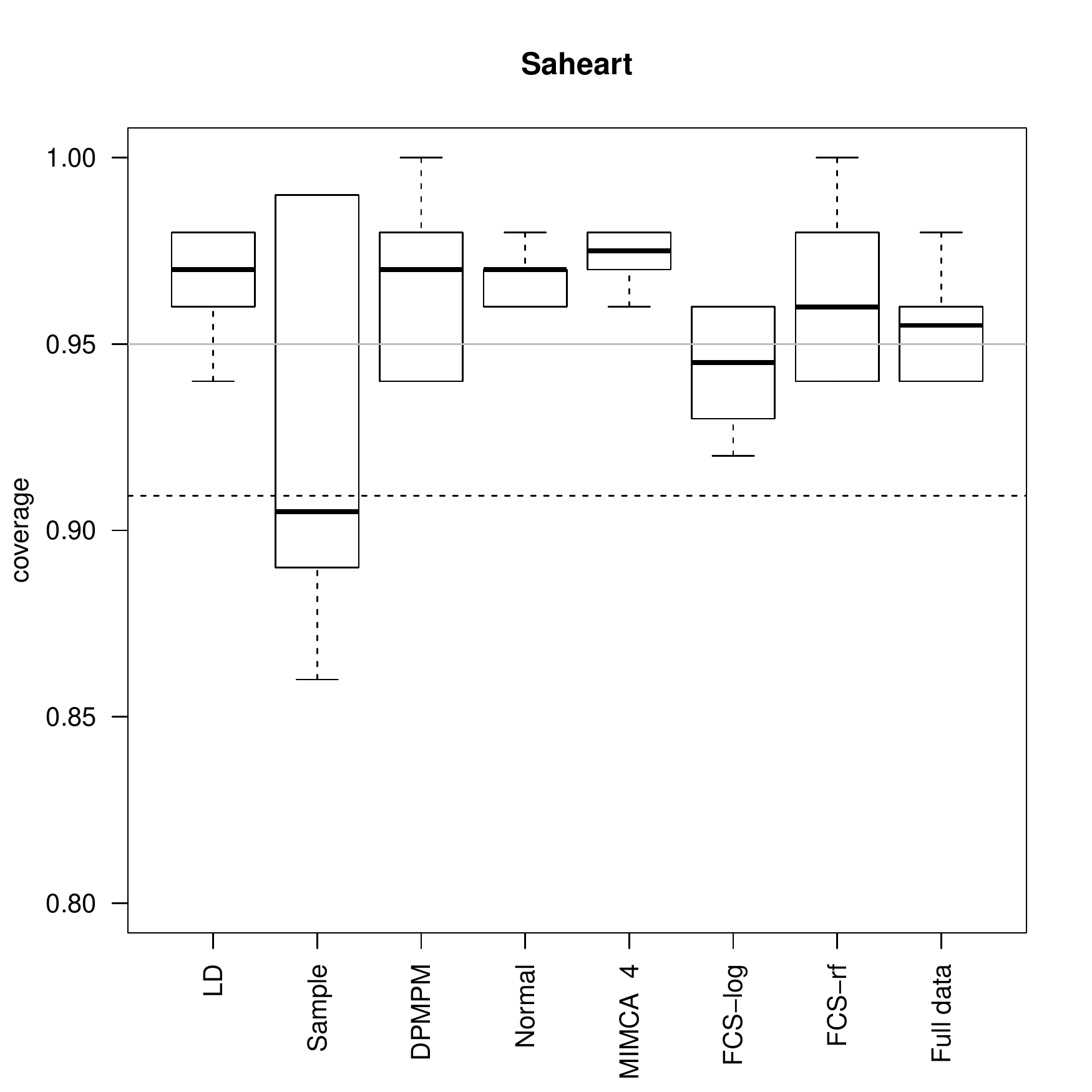}
\includegraphics[scale=.3]{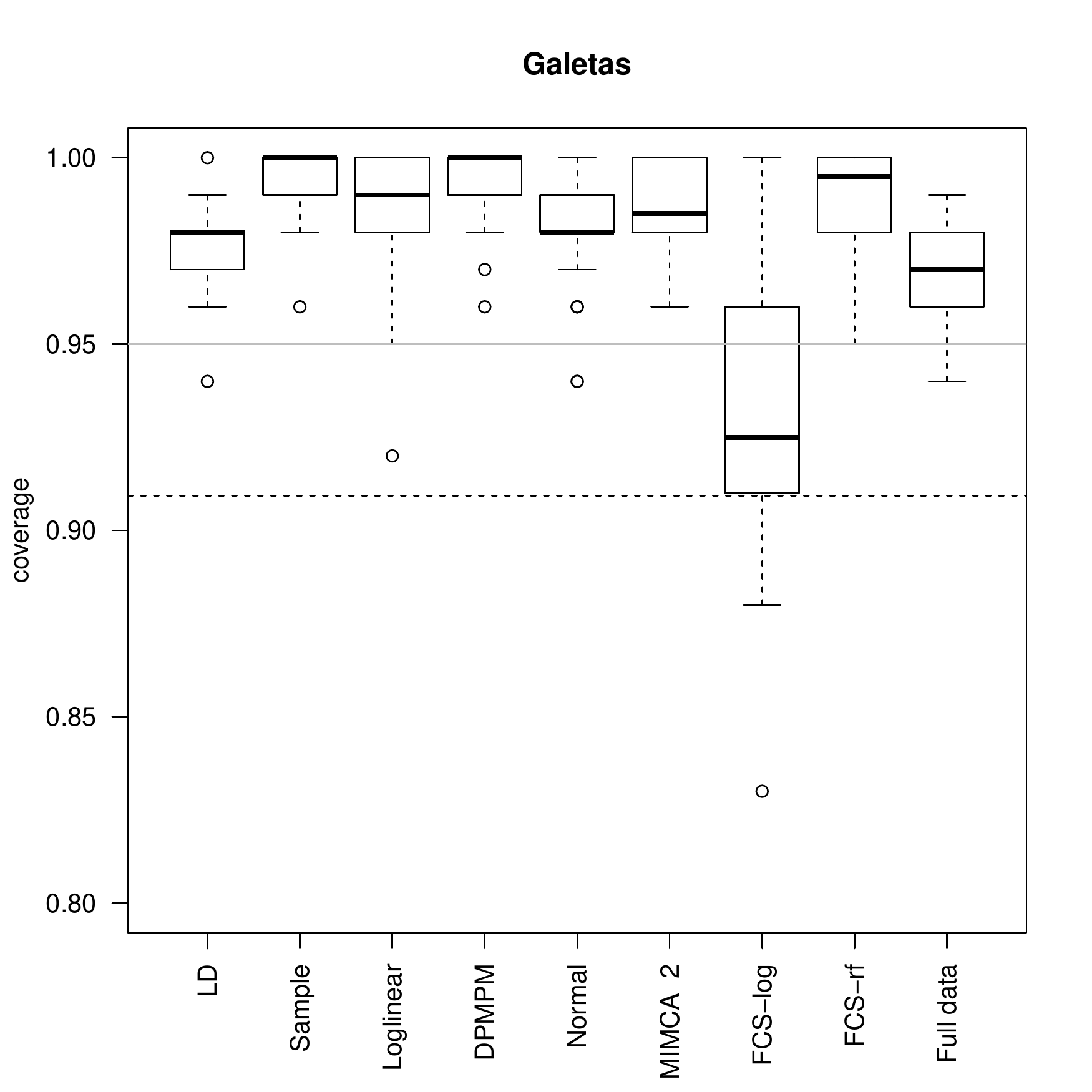}
\includegraphics[scale=.3]{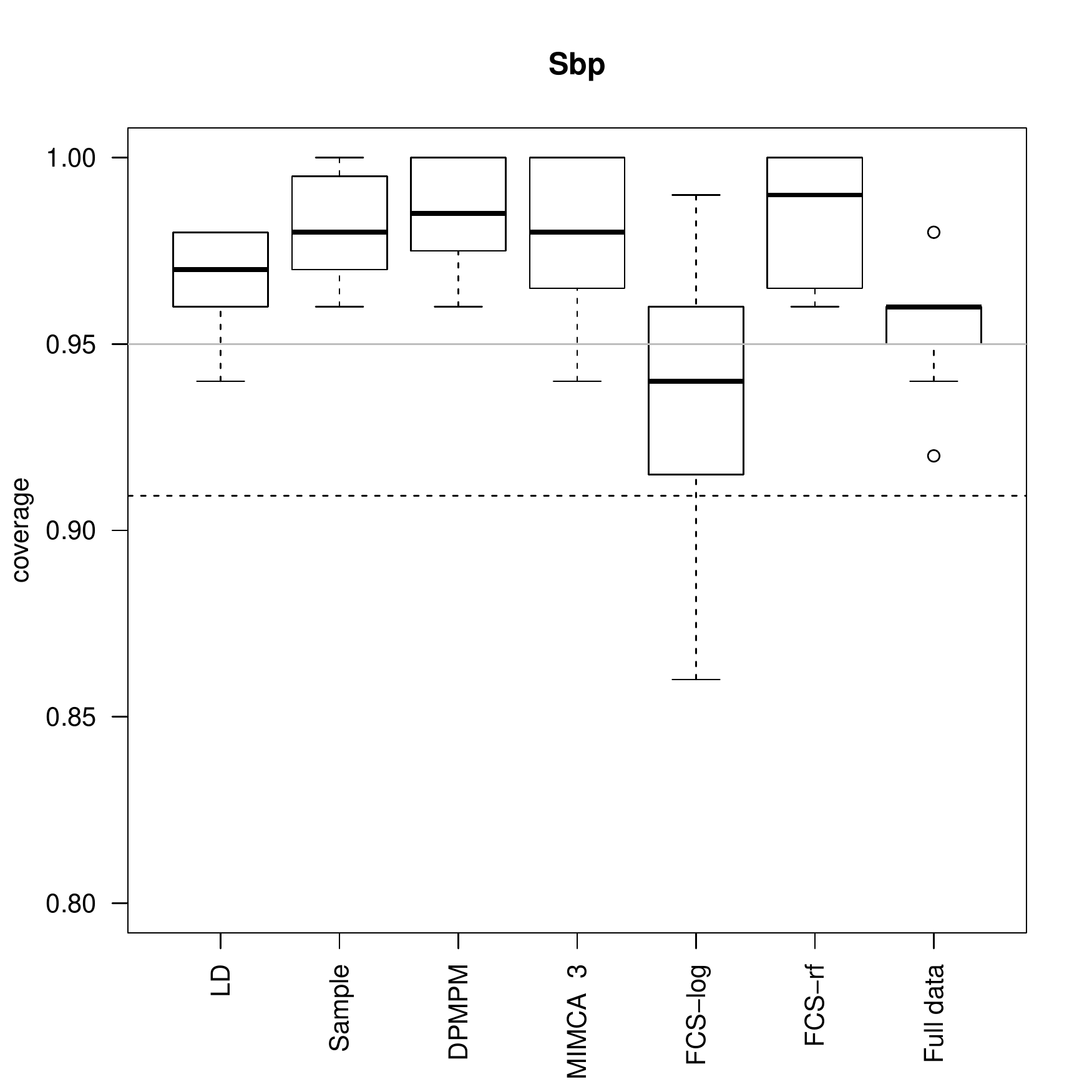}
\includegraphics[scale=.3]{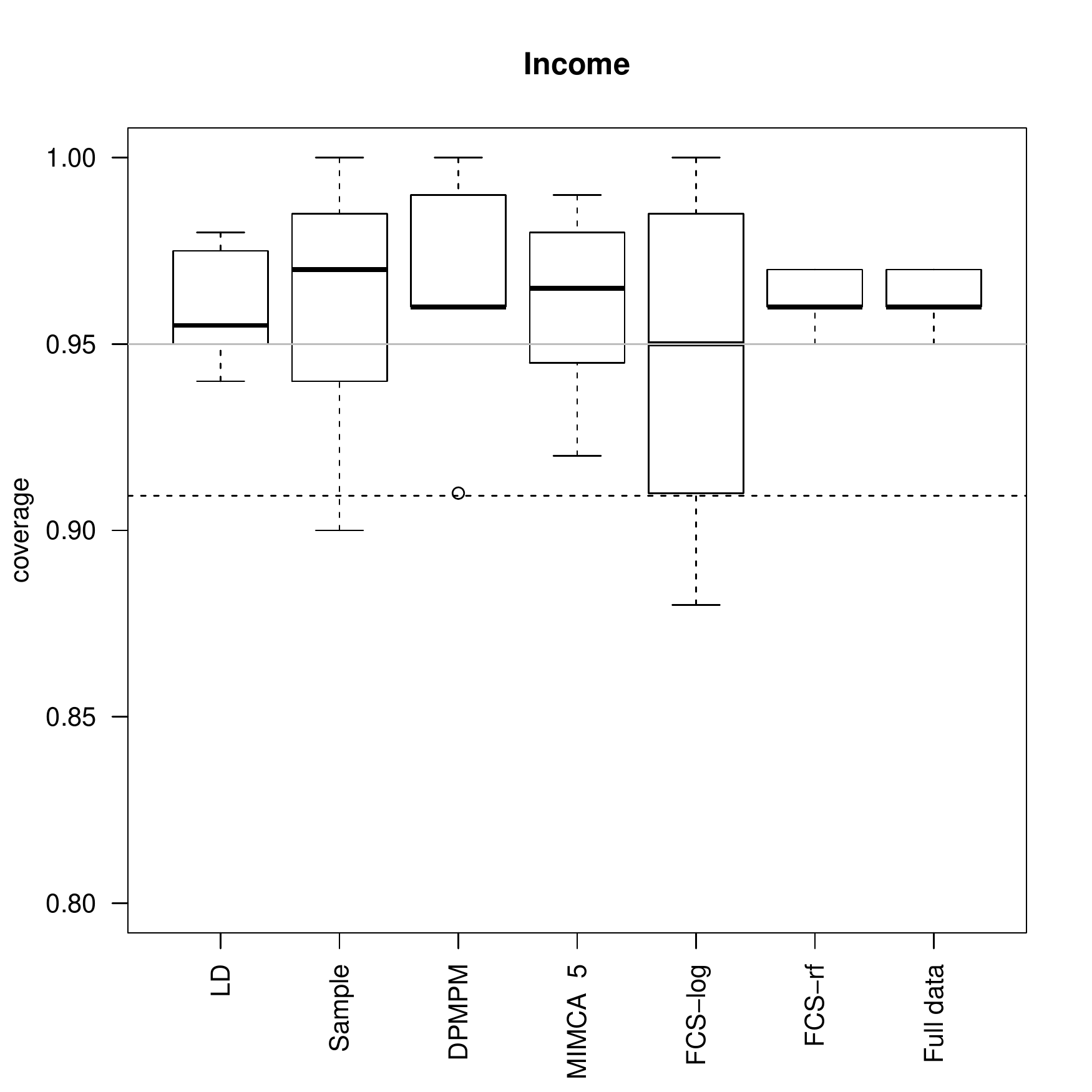}
\includegraphics[scale=.3]{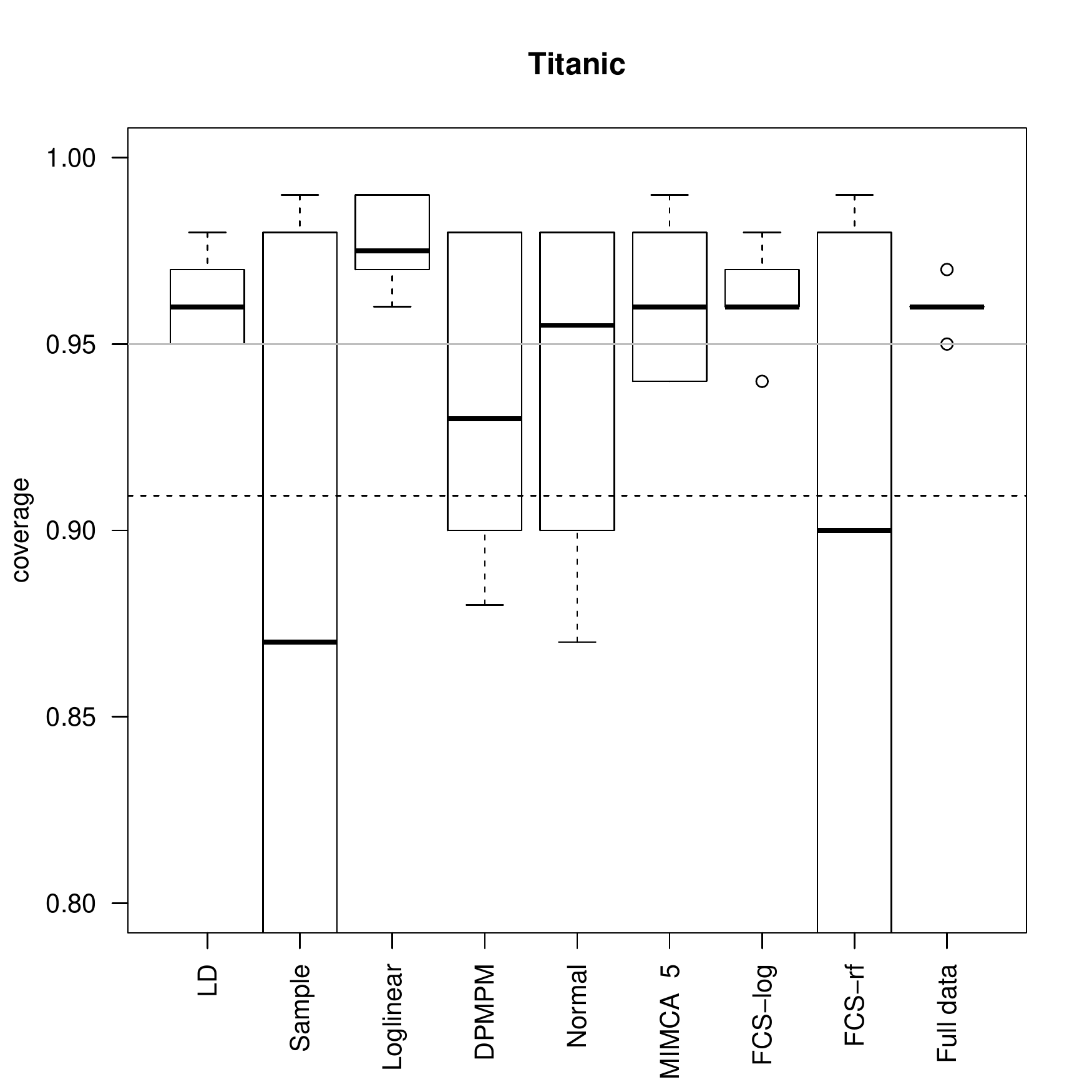}
\includegraphics[scale=.3]{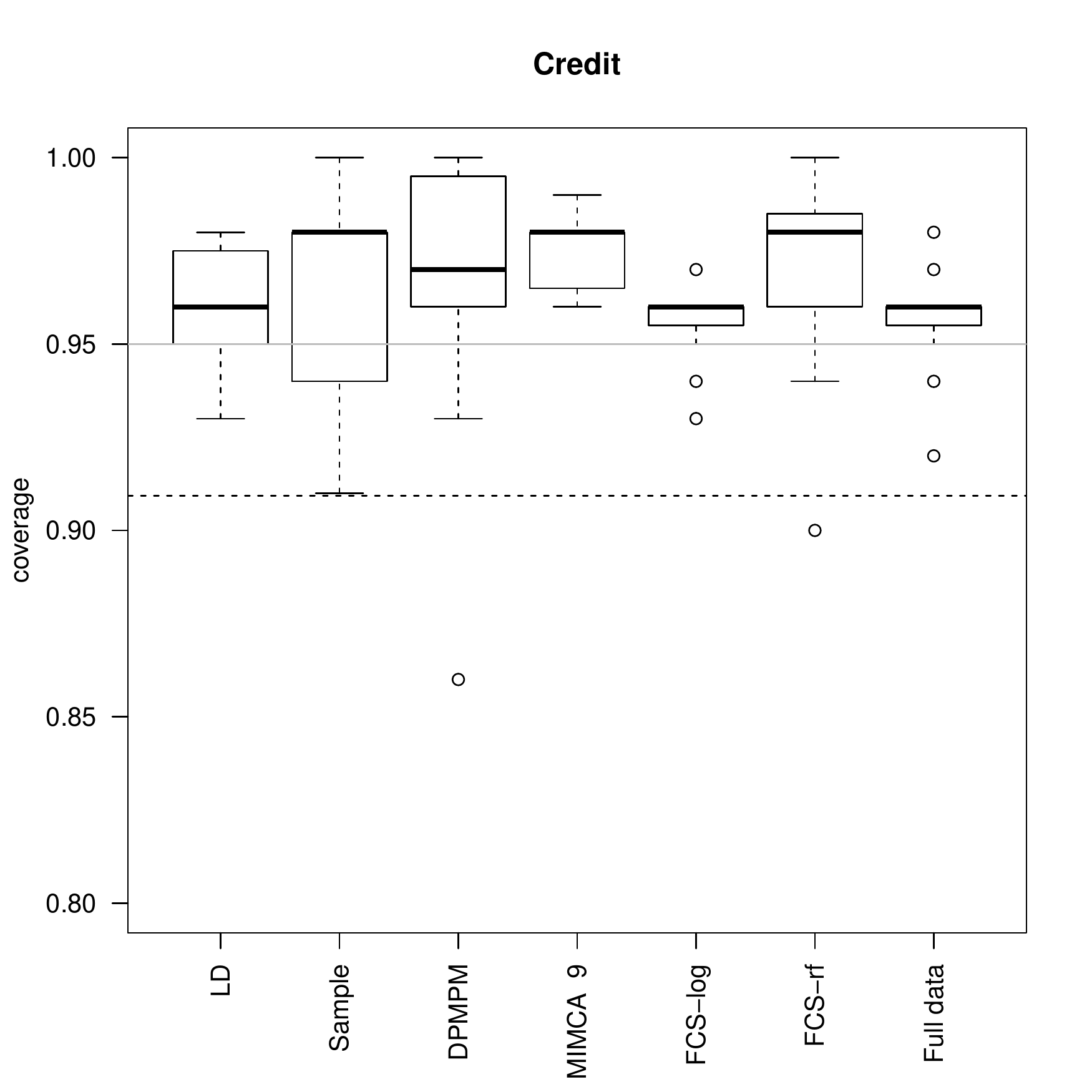}
\caption{Distribution of the coverages of the confidence intervals for all the parameters, for several methods (Listwise deletion, Sample, Loglinear model, Normal distribution, DPMPM, MIMCA, FCS using logistic regressions, FCS using random forests, Full data) and for different data sets (Saheart, Galetas, Sbp, Income, Titanic, Credit). The horizontal dashed line corresponds to the lower bound of the 95\% confidence interval for a proportion of 0.95 from a sample of size 200 according to the Agresti-Coull method \cite{AgrestiCoull}. Coverages under this line are considered as undesirable.\label{figcov}}
\end{center}
\end{figure}

For each data set and each method, the coverages of all the confidence intervals of the parameters of the model are calculated from $\Nbsim$ simulations (see Table \ref{tablepar} in Appendix \ref{modellog} for more details on these models). All the coverages are summarized with a boxplot (see Figure \ref{figcov}). The results for the bias and the confidence interval width are presented in Figure \ref{figbias} and \ref{figci} in Appendix \ref{rescomp}.

As expected, MI using the loglinear model performs well on the two data sets where it can be applied. The coverages are close to the nominal levels, the biases are close to zero, and the confidence interval widths are small.

MI using the non-parametric version of the latent class model performs quite well since most of the quantities of interest have a coverage close to 95\%. However, some inferences are incorrect from time to time such as on the data set \textit{Credit} or \textit{Titanic}. This behaviour is in agreement with the study of \cite{Si2013} which also presents some unsatisfactory coverages. \cite{Vidotto14} note that this MI model can have some difficulties in capturing the associations among the variables, particularly when the number of variables is high or the relationships between variables are complex, that can explain the poor coverages observed. Indeed, on the data set \textit{Credit}, the number of variables is the highest among the data sets considered, while on the data set \textit{Titanic}, the relationships between variables can be described as complex, in the sense that the survival status of the passengers is linked to all the other variables, but these are not closely connected. Moreover, the very poor coverages for the method Sample indicates that the imputation model has to take into account these relationships to provide confidence intervals that reach the nominal rate.

MI using the normal distribution can be applied on three data sets only. On these data sets, the coverages can be too small (see \textit{Titanic} in Figure \ref{figcov}). This highlights that despite the fact that this method is still often used in practice to deal with incomplete categorical data, it is not suitable and we do not recommend using such a strategy. However, \cite{Schafer97} showed that this method could be used to impute mixed data (\textit{i.e.} with continuous and categorical data) but only continuous variables contain missing values. 

The FCS using logistic regressions encounters difficulties on the data sets with a high number of categories such as \textit{Galetas} and \textit{Income}. This high number of categories implies a high number of parameters for each conditional model that may explain the undercoverage on several quantities.

The FCS using random forests performs well and the method encounters difficulties only on the \textit{Titanic} data set. 
This behaviour can be explained by the step of subsampling variables in the imputation algorithm (Section \ref{fcs}), \textit{i.e.}, each tree is built with potentially different variables and with a smaller number than $(\Nbvar-1)$. In the \textit{Titanic} data set, the number of variables is very small and the relationships between the variables are weak and all the variables are important to predict the survival response. Thus, it introduces too much bias in the individual tree prediction which may explain the poor inference. Even if, in the most practical cases, MI using random forests is very robust to the misspecification of the parameters, on this data set, the inference could be improved in increasing the number of explanatory variables retained for each tree.

Concerning MI using MCA, all the coverages observed are satisfying. The confidence interval width is of the same order of magnitude than the other MI methods. In addition, the method can be applied whatever the number of categories  per variables, the number of variables or the number of individuals. Thus, it appears to be the easiest method to use to impute categorical data. 

\subsubsection{Computational efficiency}
MI methods can be time consuming and the running time of the algorithms could be considered as an important property of a MI method from a practical point of view. Table \ref{tabletime} gathers the times required to impute $\Nbtab=5$ times the data sets with 20\% of missing values.
\begin{table}[h!]
\begin{center}
\begin{tabular}{|p{2.5cm}|r|r|r|r|r|r|}
\hline
        &Saheart &Galetas  &  Sbp  &Income &Titanic &Credit\\ \hline
Loglinear  &    NA &  4.597 &    NA &     NA &  0.740 &    NA\\ \hline
DPMPM    &20.050  &17.414 &56.302 &143.652 & 10.854 &24.289\\ \hline
Normal   & 0.920  & 0.822  &   NA & 26.989  & 0.483  &   NA\\ \hline
MIMCA  & 5.014   &8.972  &7.181  &58.729  & 2.750  &8.507\\ \hline
FCS log    &20.429&  38.016 &53.109& 881.188 &  4.781 &56.178\\ \hline
FCS forests & 91.474 & 112.987 & 193.156& 6329.514 & 265.771 & 461.248\\ \hline
\end{tabular}
\end{center}
\caption{Time consumed (in seconds) to impute data sets (Saheart, Galetas, Sbp, Income, Titanic, Credit), for different methods (Loglinear model, DPMPM, Normal distribution, MIMCA, FCS using logistic regressions, FCS using random forests). The imputation is done for $\Nbtab=5$ data sets. Calculation has been performed on an Intel\textregistered  Core\texttrademark 2 Duo CPU E7500, running Ubuntu 12.04 LTS equipped with 3 GB ram. Some values are not provided because all methods cannot be performed on each data set.\label{tabletime}}
\end{table}

First of all, as expected, the FCS method is more time consuming than the others based on a joint model. In particular, for the data set \textit{Income}, where the number of individuals and variables is high, the FCS using random forests requires 6,329 seconds (\textit{i.e.} 1.75 hours), illustrating substantial running time issues. FCS using logistic regressions requires 881 seconds, a time 6 times higher than MI using the latent class model, and 15 times higher than  MI method using MCA. Indeed, the number of incomplete variables increases the number of conditional models required, as well as the number of parameters in each of them because more covariates are used. In addition, the time required to estimate its parameters is non-negligible, particularly when the number of individuals is high. Then, MI using the latent class model can be performed in a reasonable time, but this is at least two times higher than the one required for MI using MCA. Thus, the MIMCA method should be particularly recommended to impute data sets of high dimensions.

Having a method which is not too expensive enables the user to produce more than the classical $\Nbtab=5$ imputed data sets. This could lead to a more accurate inference. 

\subsubsection{Choice of the number of dimensions\label{nbaxe}}


MCA requires a predefined number of dimensions $\Nbdim$ which can be chosen by cross-validation \cite{Josse12}. Cross-validation consists in searching the number of dimensions $\Nbdim$ minimizing an error of prediction. More precisely, missing values are added completely at random to the data set $\bfX$. Then, the missing values of the incomplete disjunctive table $\bfZ$ are predicted using the regularized iterative MCA algorithm. The mean error of prediction is calculated according to $\frac{1}{Card(\mathcal{U})}\sum_{(\nbind,\nbcol)\in \mathcal{U}}{(z_{\nbind \nbcol}-\hat z_{\nbind \nbcol})^2}$, where $\mathcal{U}$ denotes the set of the added missing values. The procedure is repeated $k$ times for a predefined number of dimensions. The number of dimensions retained is the one minimizing the mean of the $k$ mean errors of prediction. This procedure can be used whether the data set contains missing values or not.

\begin{figure}[h!]
\begin{center}
\includegraphics[scale=.3]{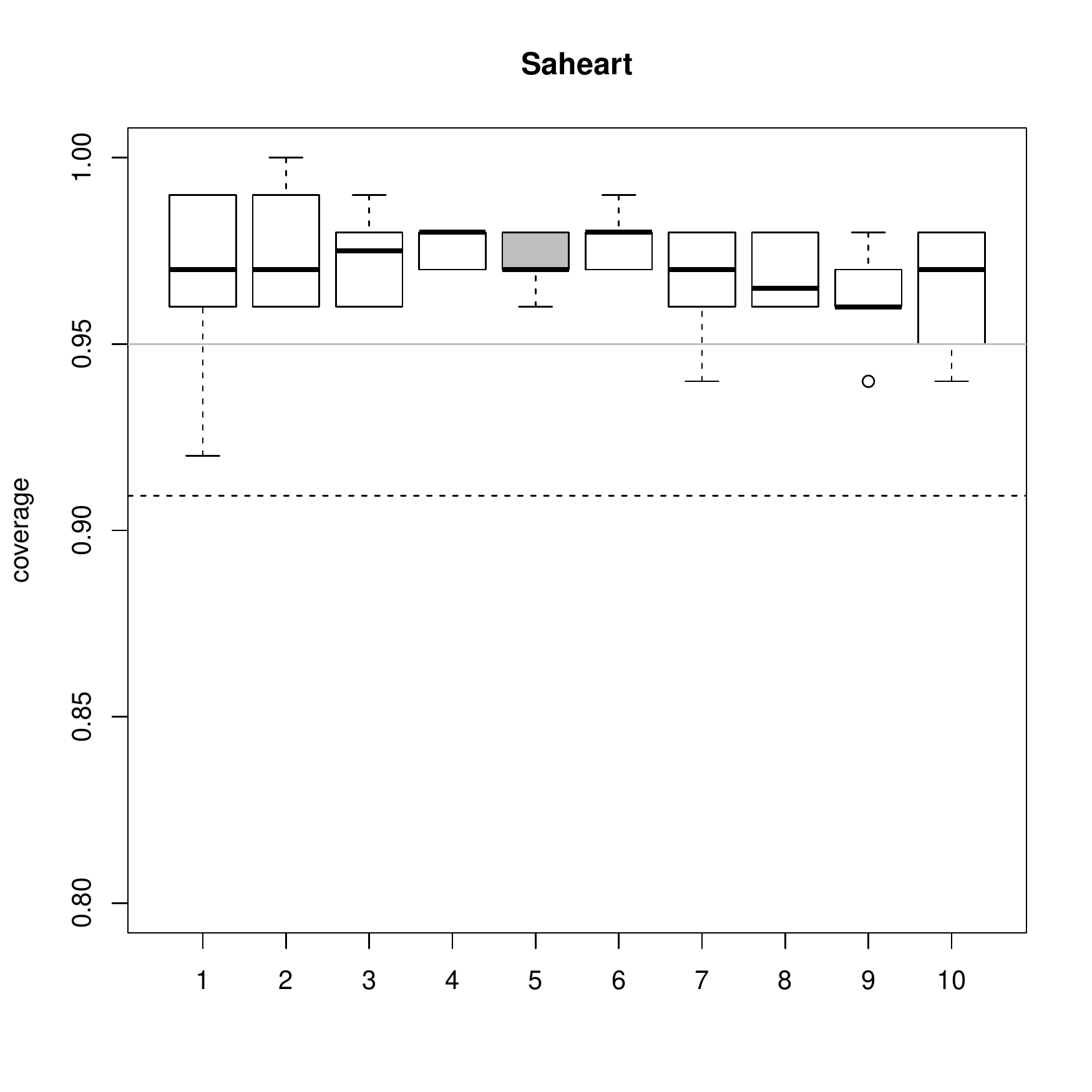}
\includegraphics[scale=.3]{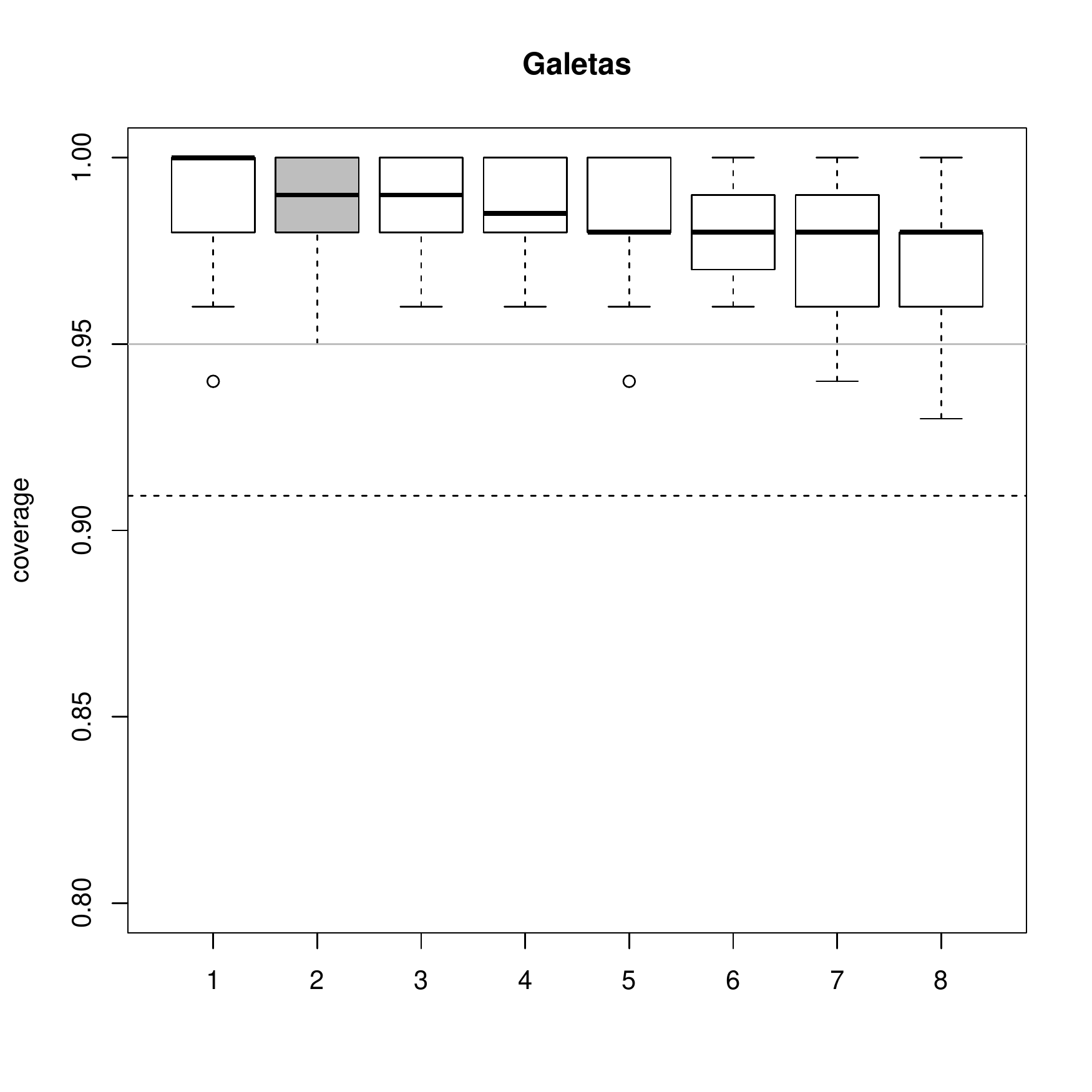}
\includegraphics[scale=.3]{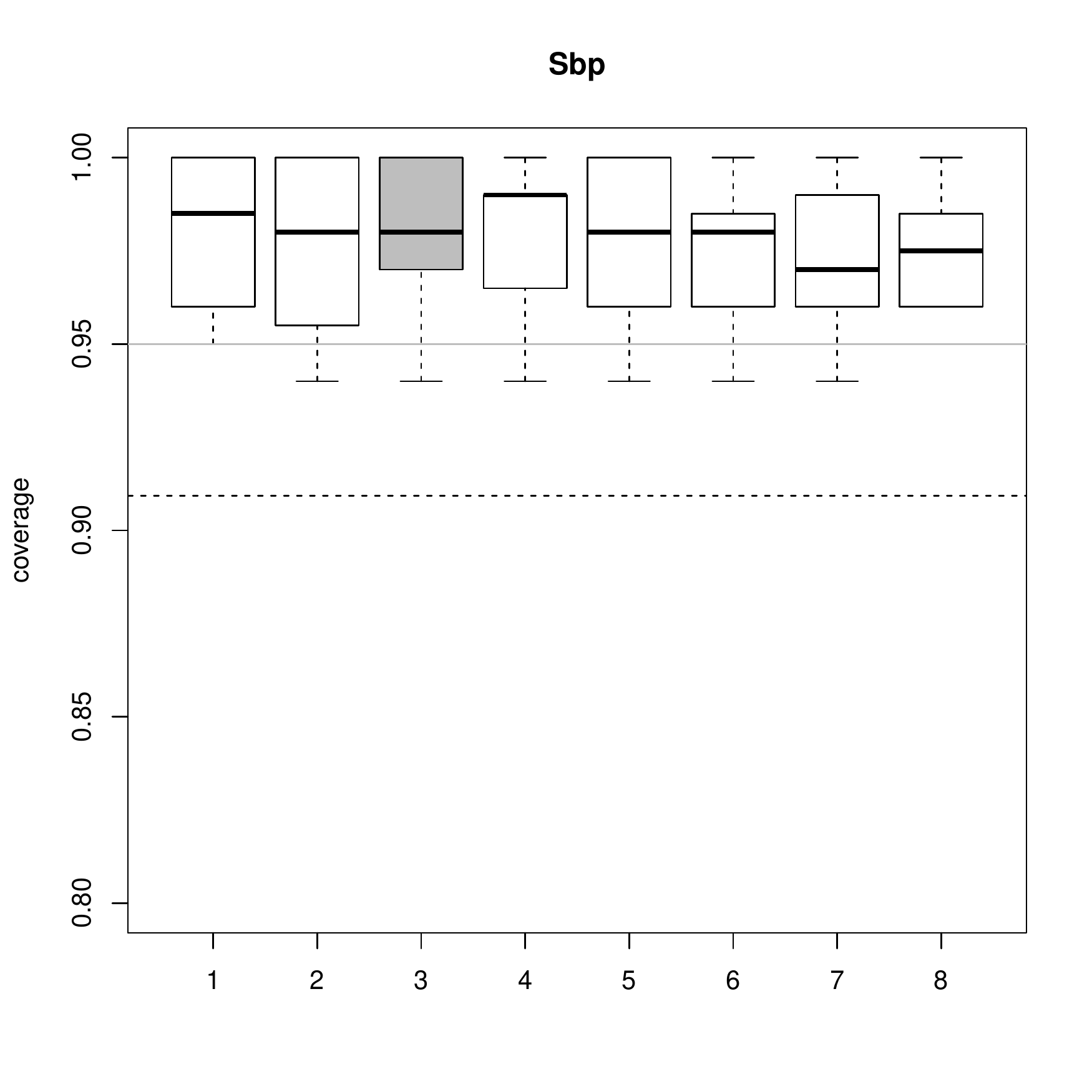}
\includegraphics[scale=.3]{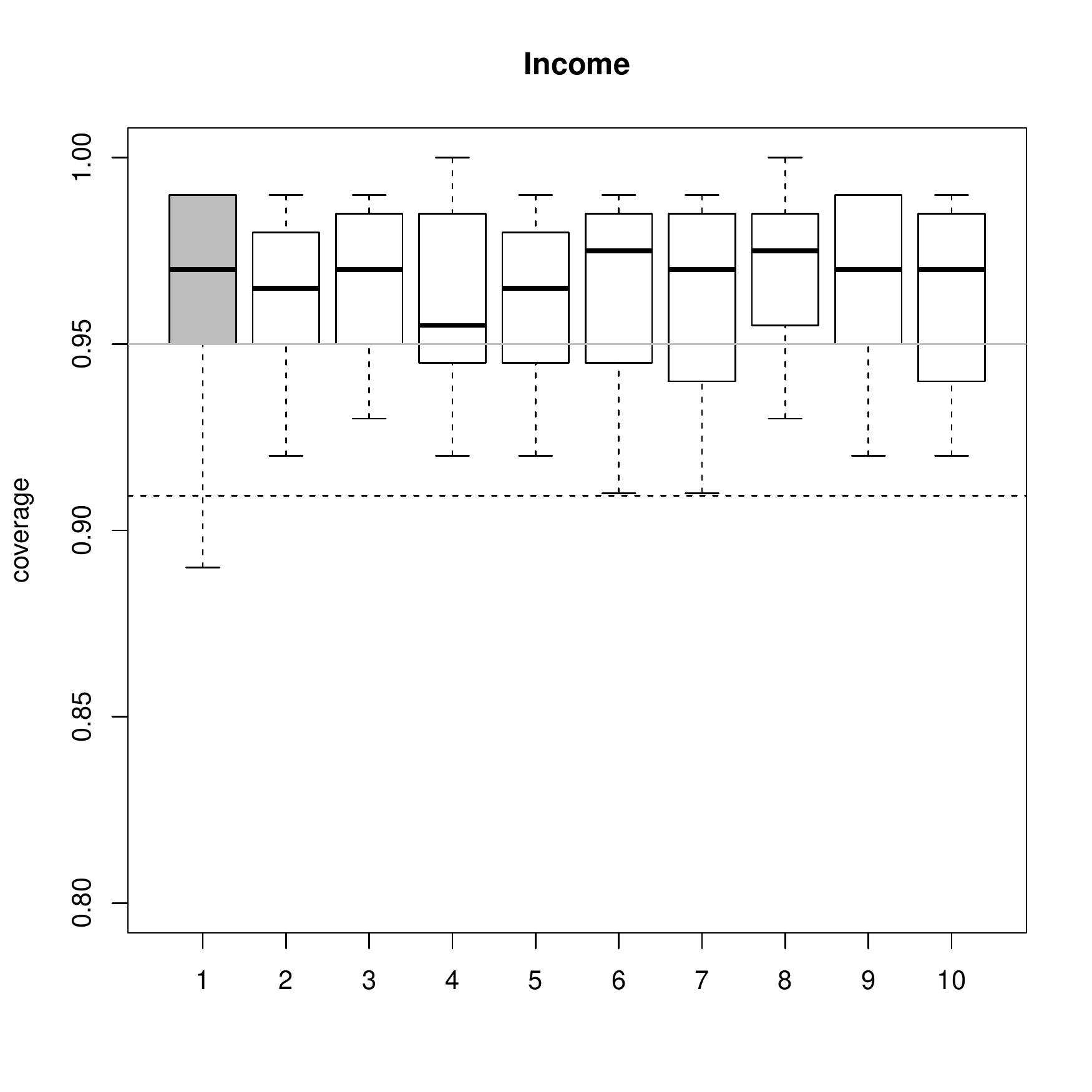}
\includegraphics[scale=.3]{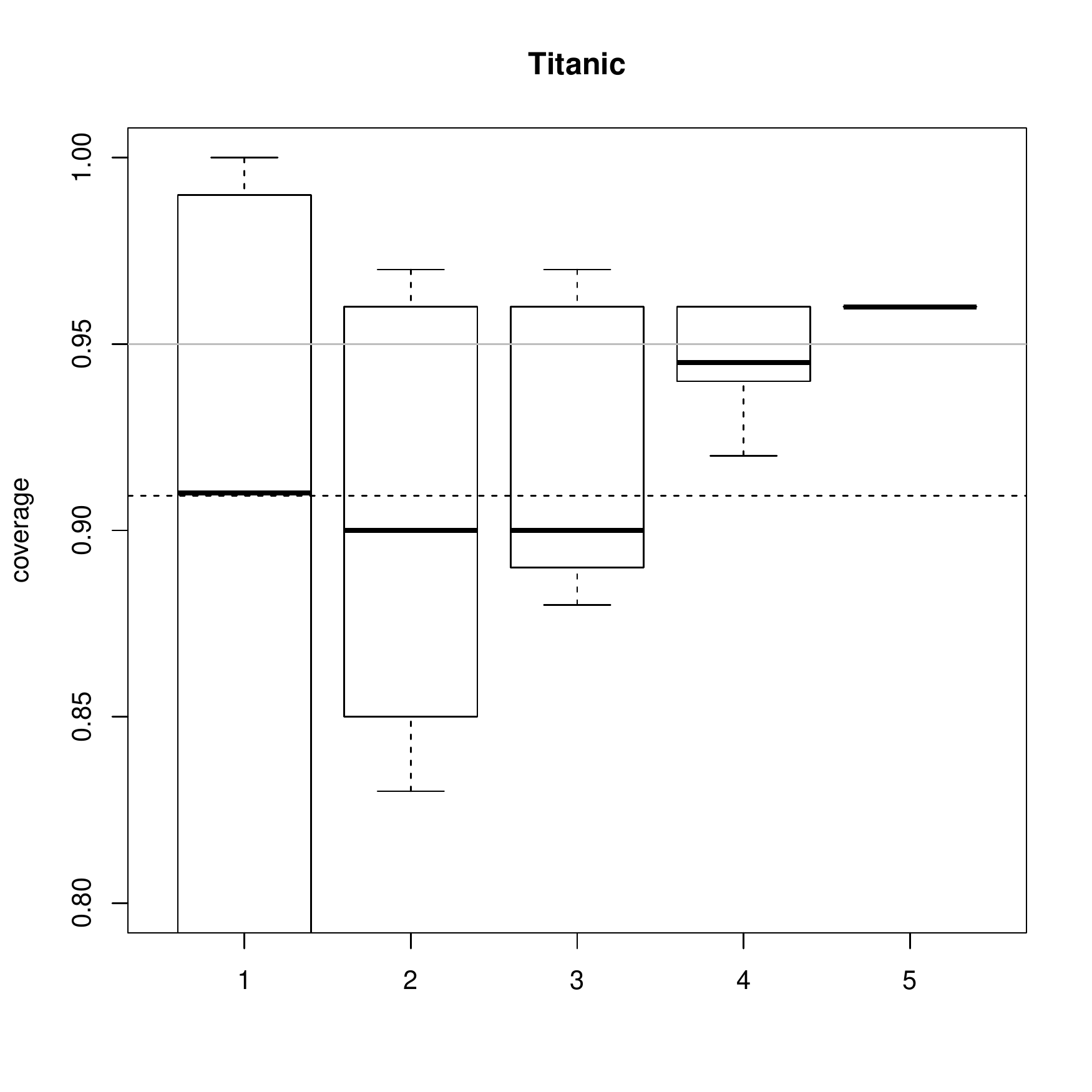}
\includegraphics[scale=.3]{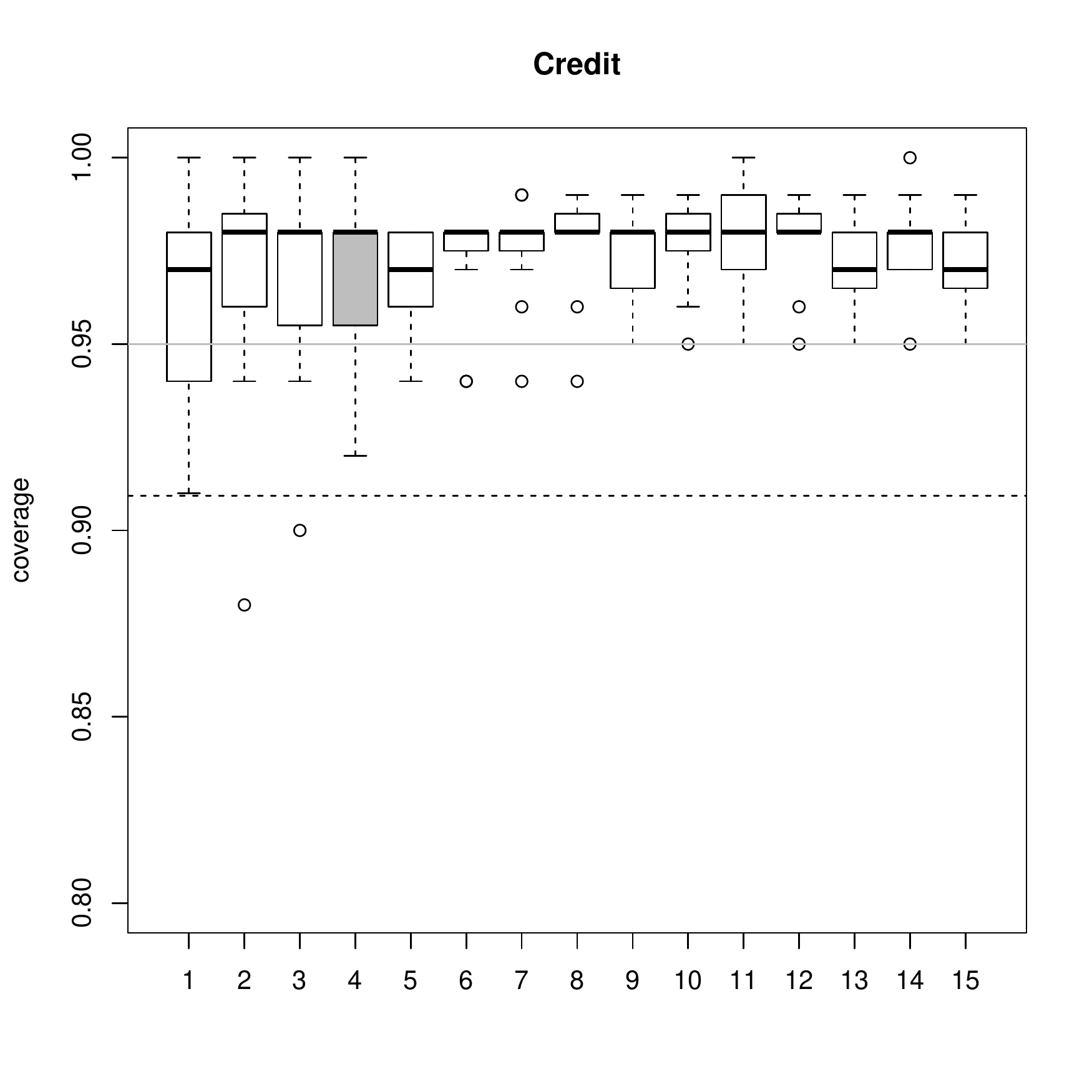}
\caption{Distribution of the coverages of the confidence intervals for all the parameters for the MIMCA algorithm for several numbers of dimensions and for different data sets (Saheart, Galetas, Sbp, Income, Titanic, Credit). The results for the number of dimensions provided by cross-validation are in grey. The horizontal dashed line corresponds to the lower bound of the 95\% confidence interval for a proportion of 0.95 from a sample of size 200 according to the Agresti-Coull method \cite{AgrestiCoull}. Coverages under this line are considered as undesirable. \label{fignbaxecov}}
\end{center}
\end{figure}

To evaluate how the choice of $\Nbdim$ impacts on the quality of the inferences, we perform the MIMCA algorithm varying the number of dimensions around the one provided by cross-validation. Figure \ref{fignbaxecov} presents how this tuning parameter influences the coverages in the previous study. The impacts on the width of the confidence intervals are reported in Figure \ref{fignbaxeci} and the ones on the bias in Figure \ref{fignbaxebias} in Appendix \ref{rescomp}.

Except for the data set \textit{Titanic}, the coverages are stable according to the number of dimensions retained. In particular, the number of dimensions suggested by cross-validation provides coverages close to the nominal level of the confidence interval. In the case of the data set \textit{Titanic}, the cross-validation suggests retaining 5 dimensions, which is the choice giving the smallest confidence intervals, while giving coverages close to 95\%. But retaining less dimensions leads to worse performances since the covariates are not closely related (Section \ref{assess}). Indeed, these covariates can not be well represented within a space of low dimensions. Consequently, a high number of dimensions is required to reflect the useful associations to impute the data. \textit{Titanic} illustrates that underfitting can be problematic. The same comment is made by \cite{Vermunt08} who advise choosing a number of classes sufficiently high in the case of MI using the latent class model. However, overfitting is less problematic because it increases the variance, but it does not skip the useful information.

\section{Conclusion}
This paper proposes an original MI method to deal with categorical data based on MCA. The principal components and the loadings that are the parameters of the MCA enables the imputation of data. To perform MI, the uncertainty on these parameters is reflected using a non-parametric bootstrap, which results in a specific weighting for the individuals.

From a simulation study based on real data sets, this MI method has been compared to the other main available MI methods for categorical variables. We highlighted the competitiveness of the MIMCA method to provide valid inferences for an analysis model requiring two-way associations (such as logistic regression without interaction, or a homogeneous loglinear model, proportion, odds ratios, etc).

We showed that MIMCA can be applied to various configurations of data. In particular, the method is accurate for a large number of variables, for a large number of categories per variables and when the number of individuals is small. Moreover, the MIMCA algorithm performs fairly quickly, allowing the user to generate more imputed data sets and therefore to obtain more accurate inferences ($\Nbtab$ between 20 and 100 can be beneficial \cite[p.49]{VB12}). Thus, MIMCA is very suitable to impute data sets of high dimensions that require more computation. Note that MIMCA depends on a tuning parameter (the number of components), but we highlighted that the performances of the MI method are robust to a misspecification of it.\\




Because of the intrinsic properties of MCA, MI using MCA is appropriate when the analysis model contains two-way associations between variables such as logistic regression without interaction. To consider the case with interactions, one solution could be to introduce to the data set additional variables corresponding to the interactions. However, the new variable "interaction" is considered as a  variable in itself without taking into account its explicit link with the associated variables. It may lead to imputed values which are not in agreement with each others. This topic is a subject of intensive research for continuous variables \cite{Seaman12,Bartlett13}.

In addition, the encouraging results of the MIMCA to impute categorical data prompt the extension of the method to impute mixed data. The first research in this direction \cite{Audigier14} has shown that the principal components method dedicated to mixed data (called Factorial Analysis for Mixed Data) is efficient to perform single imputation, but the extension to a MI method requires further research.

\newpage
\bibliographystyle{unsrt}
\bibliography{biblioarticle}
\newpage
\appendix
\section*{Appendix}
\section{Simulation design: analysis models and sample characteristics\label{modellog}}
\begin{table}[h!]
\begin{tabular}{|p{2cm}|p{2cm}|p{2cm}|p{2cm}|p{5cm}|p{2cm}|}
\hline
Data set&number of individuals&number of variables&sample size&logistic regression model&number of quantities of interest\\ \hline
Saheart&462&10&300 &\textsc{chd} \verb|=| \textsc{famhist + tobacco + alcohol}&30\\ \hline
Galetas&1192&4&300&\textsc{galle} \verb|=| \textsc{grupo}&6 \\ \hline
Sbp&500&18&200&\textsc{sbp} \verb|=| \textsc{smoke + exercise + alcohol}&12\\ \hline
Income&6876&14&1500&\textsc{income} \verb|=| \textsc{sex}&8\\ \hline
Titanic&2201&4&300&\textsc{surv} \verb|=|\textsc{ class+age+sex}&6\\ \hline
Credit&982&20&300&\textsc{class} \verb|=| \textsc{checking}\_\textsc{status + duration + credit}\_\textsc{history + purpose}&11\\ \hline
\end{tabular}
\caption{Set of the sample characteristics and of the analysis models used to perform the simulation study (Section \ref{secsimu}) for the several data sets (Saheart, Galetas, Sbp, Income, Titanic, Credit).\label{tablepar}}
\end{table}
\section{Simluation study: complementary results\label{rescomp}}
\begin{figure}[h!]
\begin{center}
\includegraphics[scale=.3]{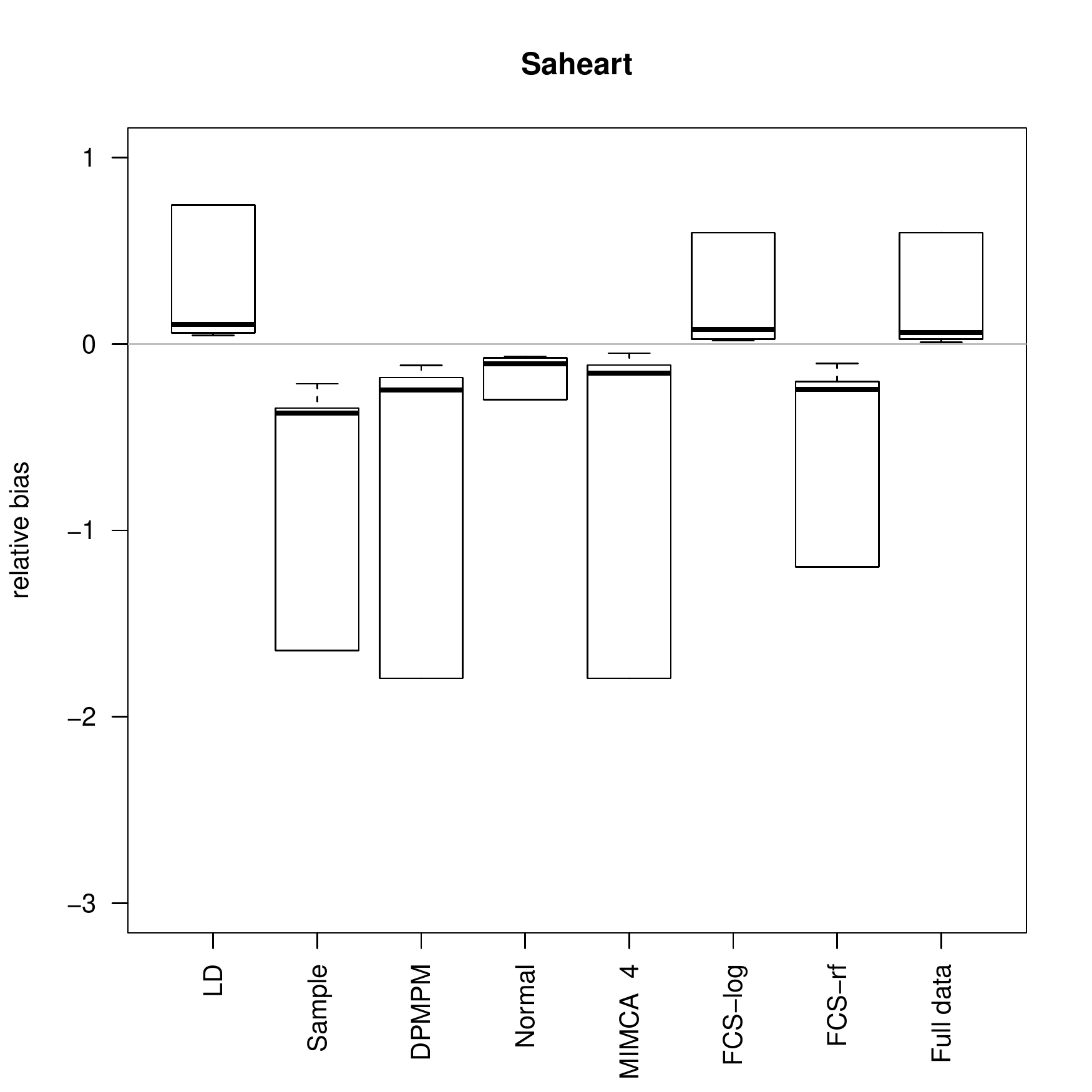}
\includegraphics[scale=.3]{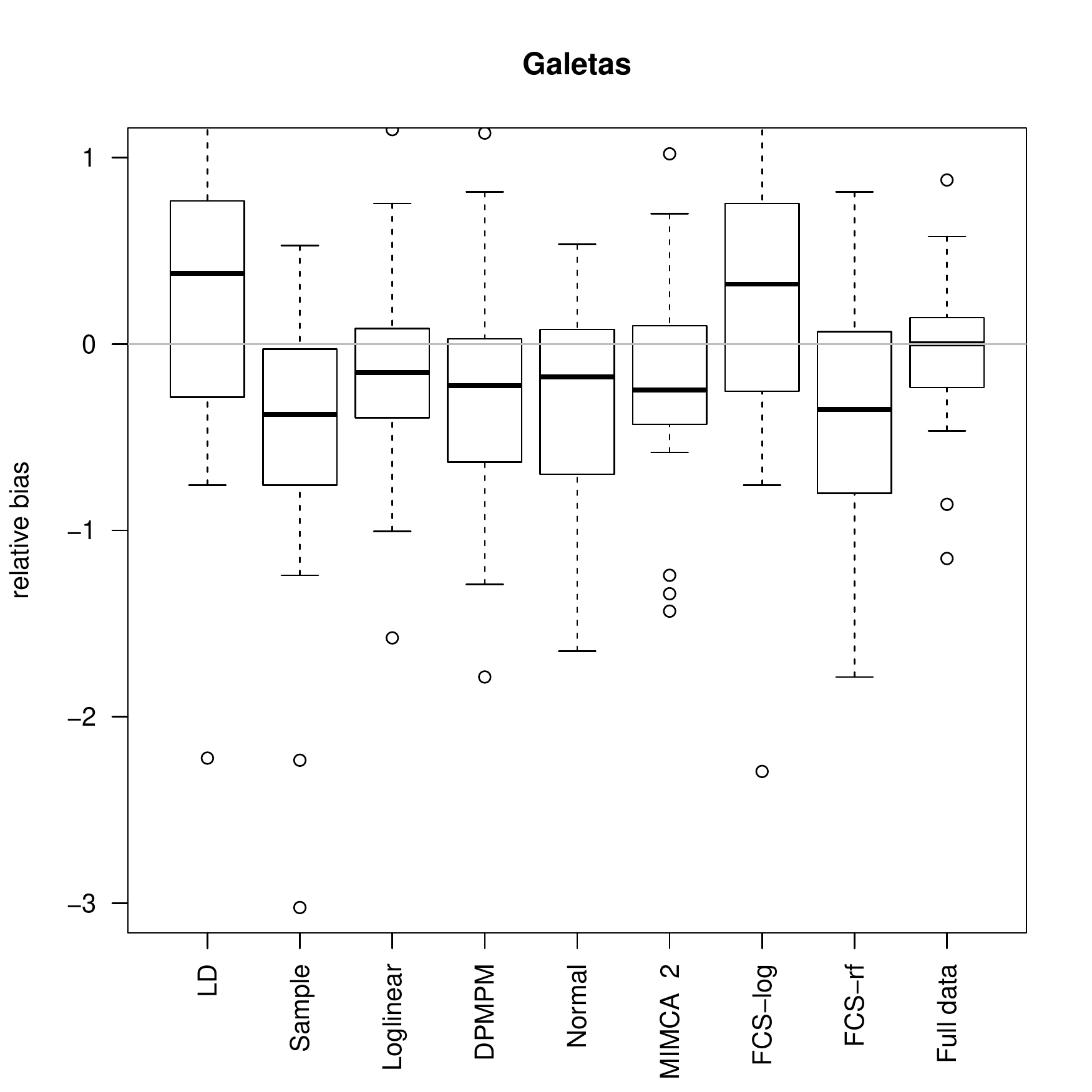}
\includegraphics[scale=.3]{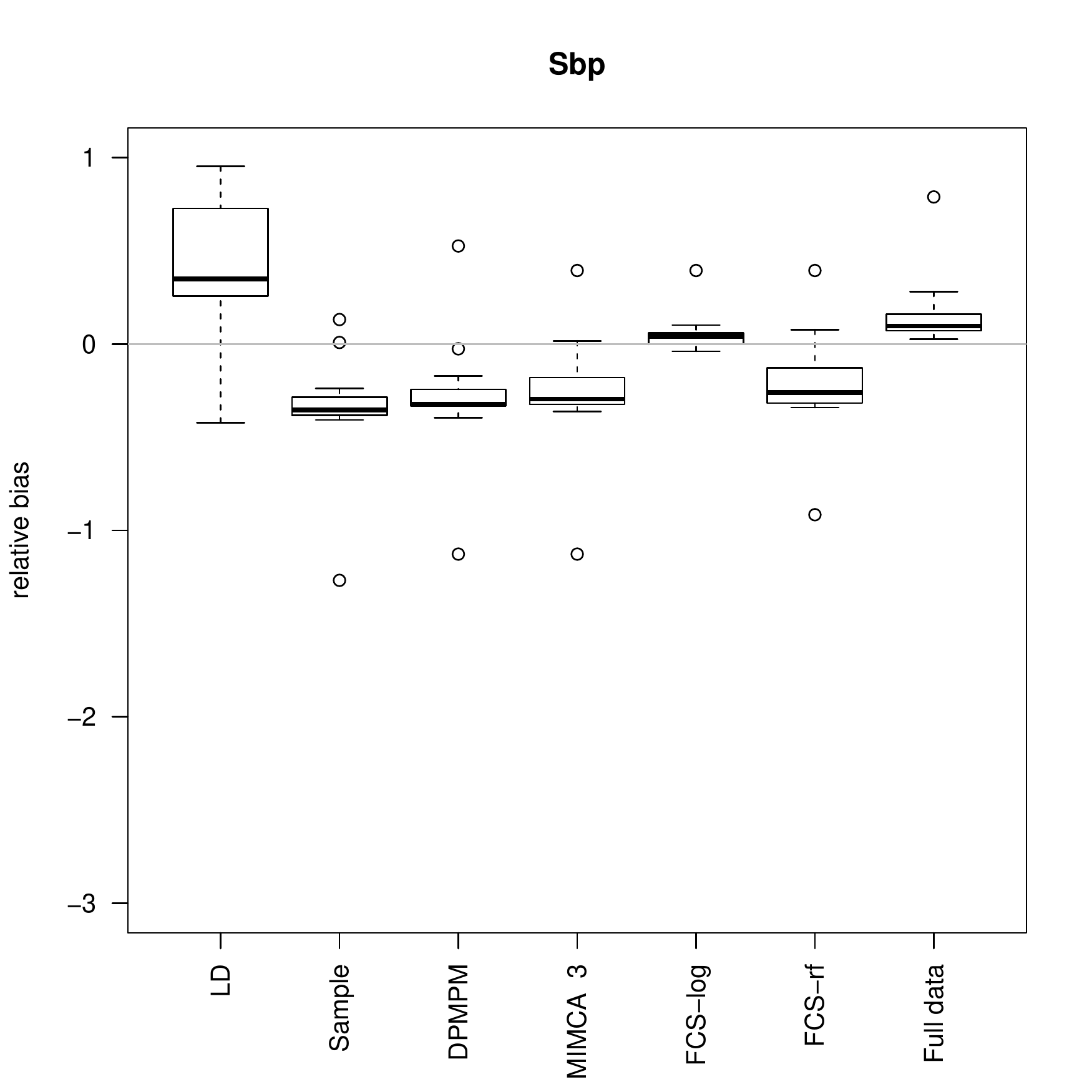}
\includegraphics[scale=.3]{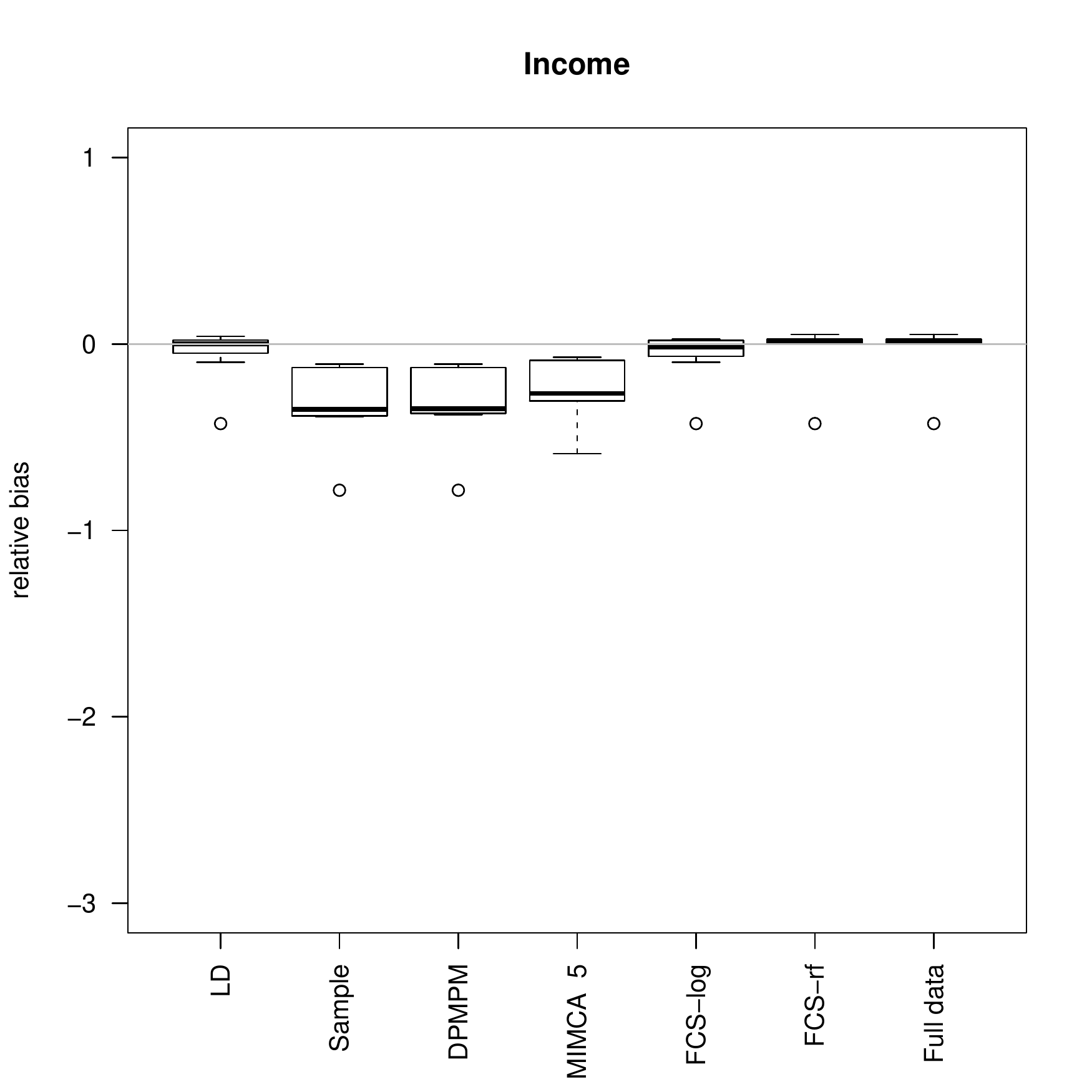}
\includegraphics[scale=.3]{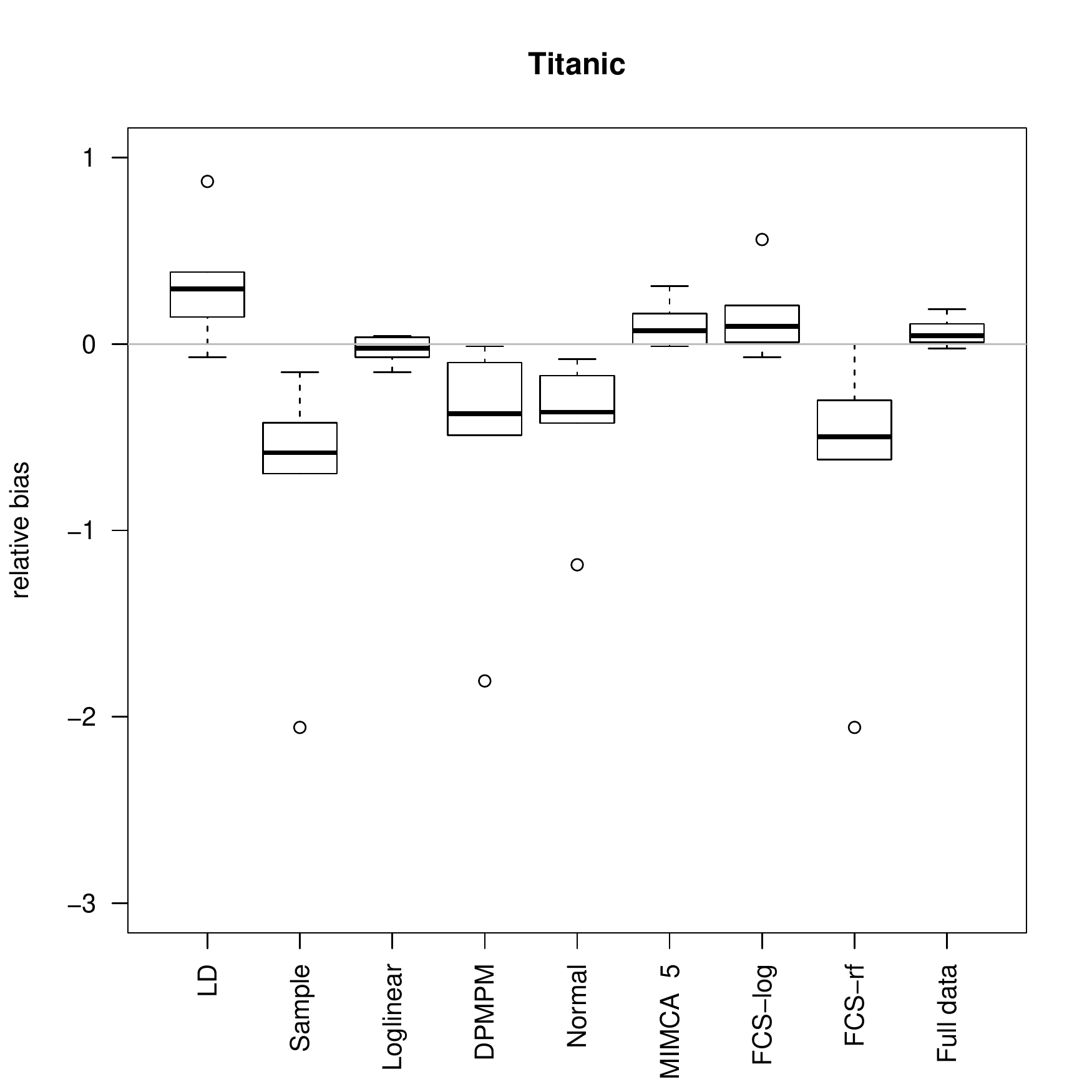}
\includegraphics[scale=.3]{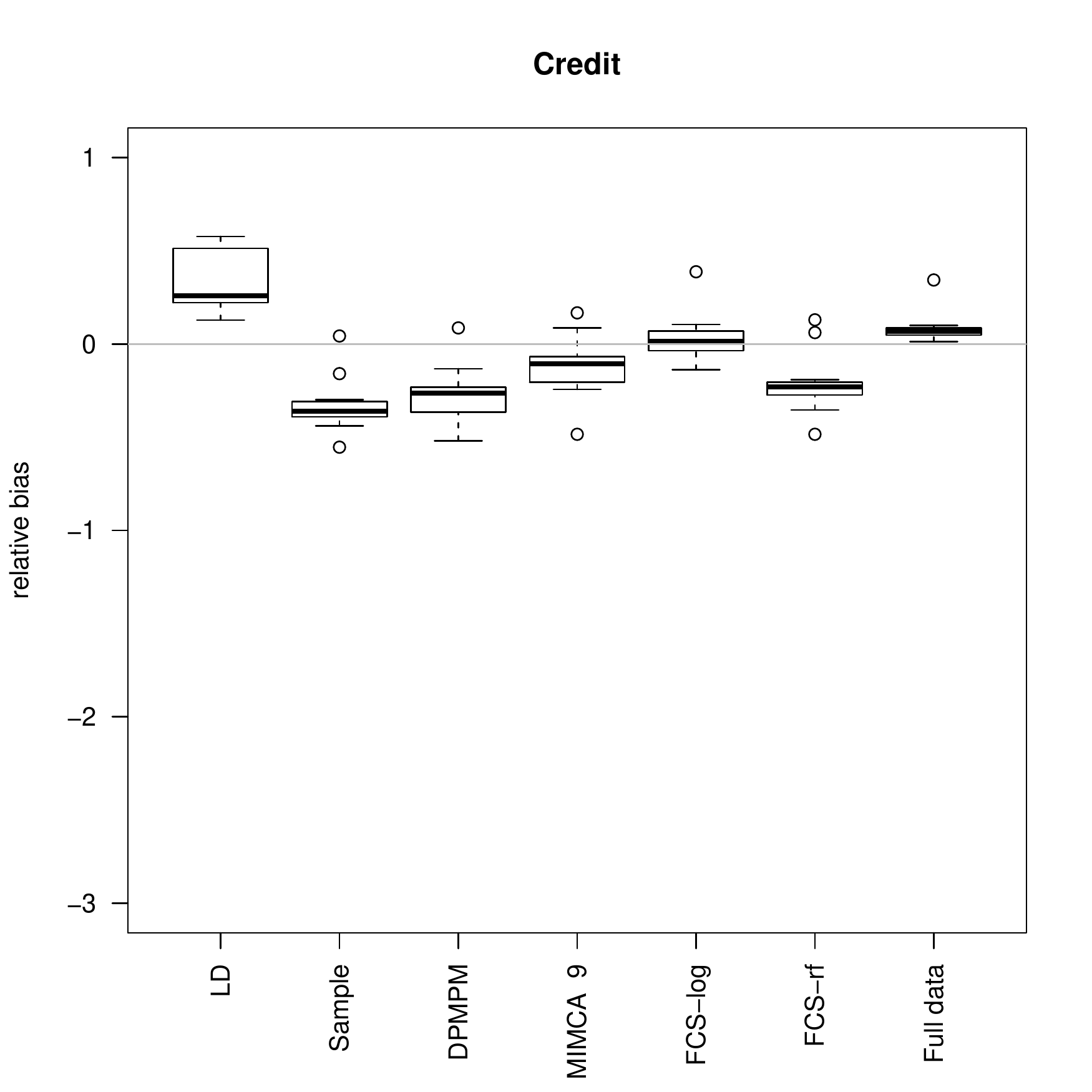}
\caption{Distribution of the relative bias (bias divided by the true value) over the several quantities of interest for several methods (Listwise deletion, Sample, Loglinear model, DPMPM, Normal distribution, MIMCA, FCS using logistic regressions, FCS using random forests, Full data) for different data sets (Saheart, Galetas, Sbp, Income, Titanic, Credit). One point represents the relative bias observed for one coefficient.\label{figbias}}
\end{center}
\end{figure}
\begin{figure}[h!]
\begin{center}
\includegraphics[scale=.3]{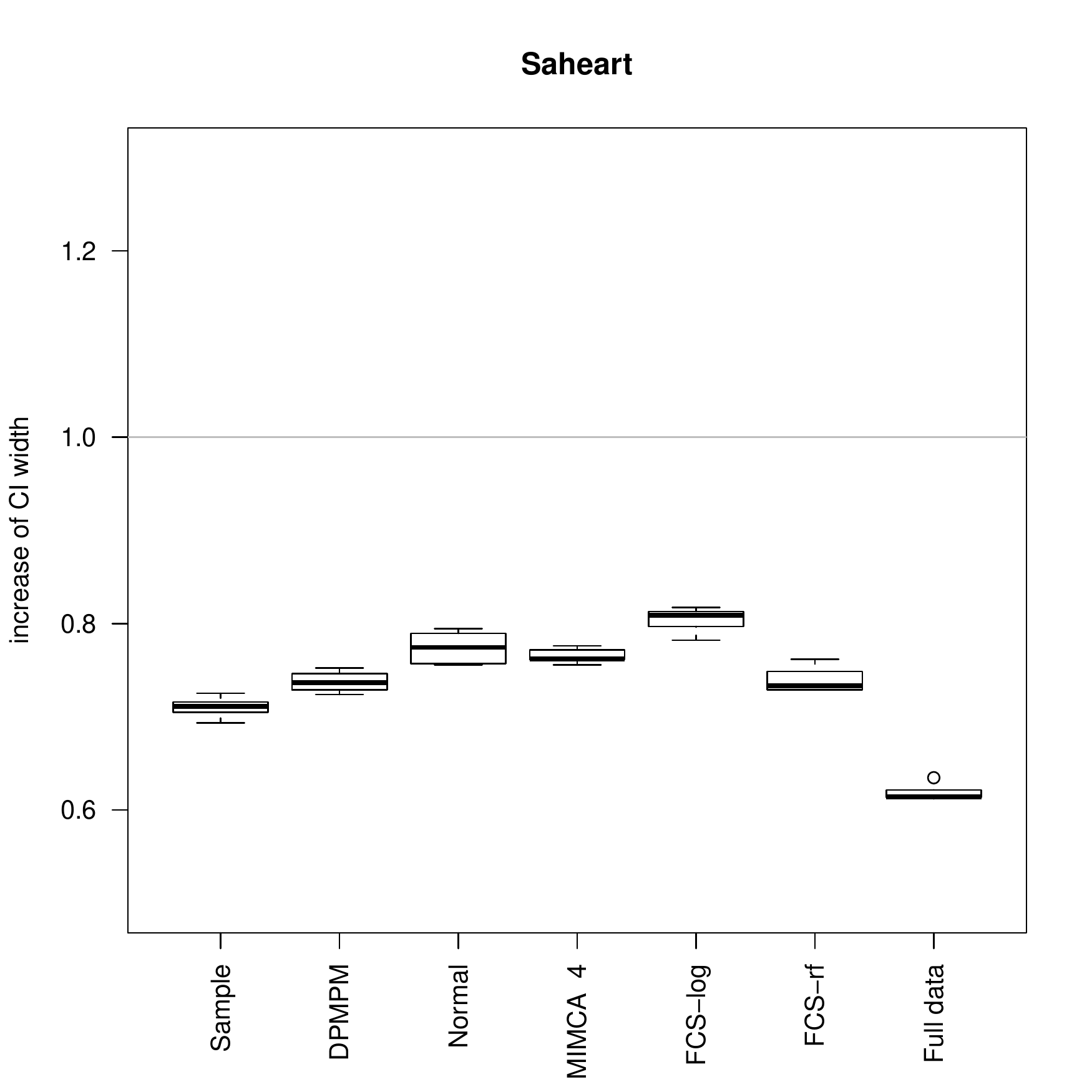}
\includegraphics[scale=.3]{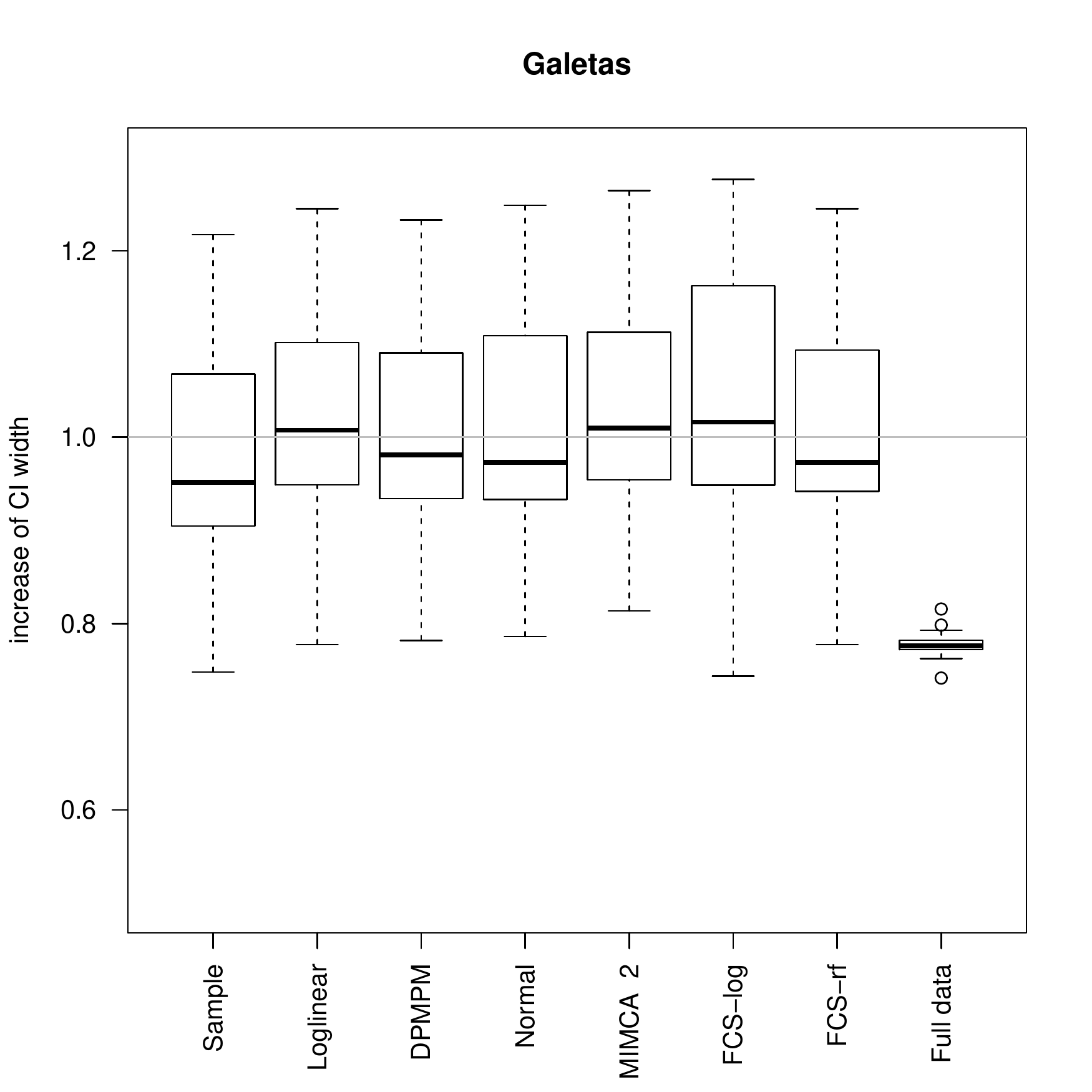}
\includegraphics[scale=.3]{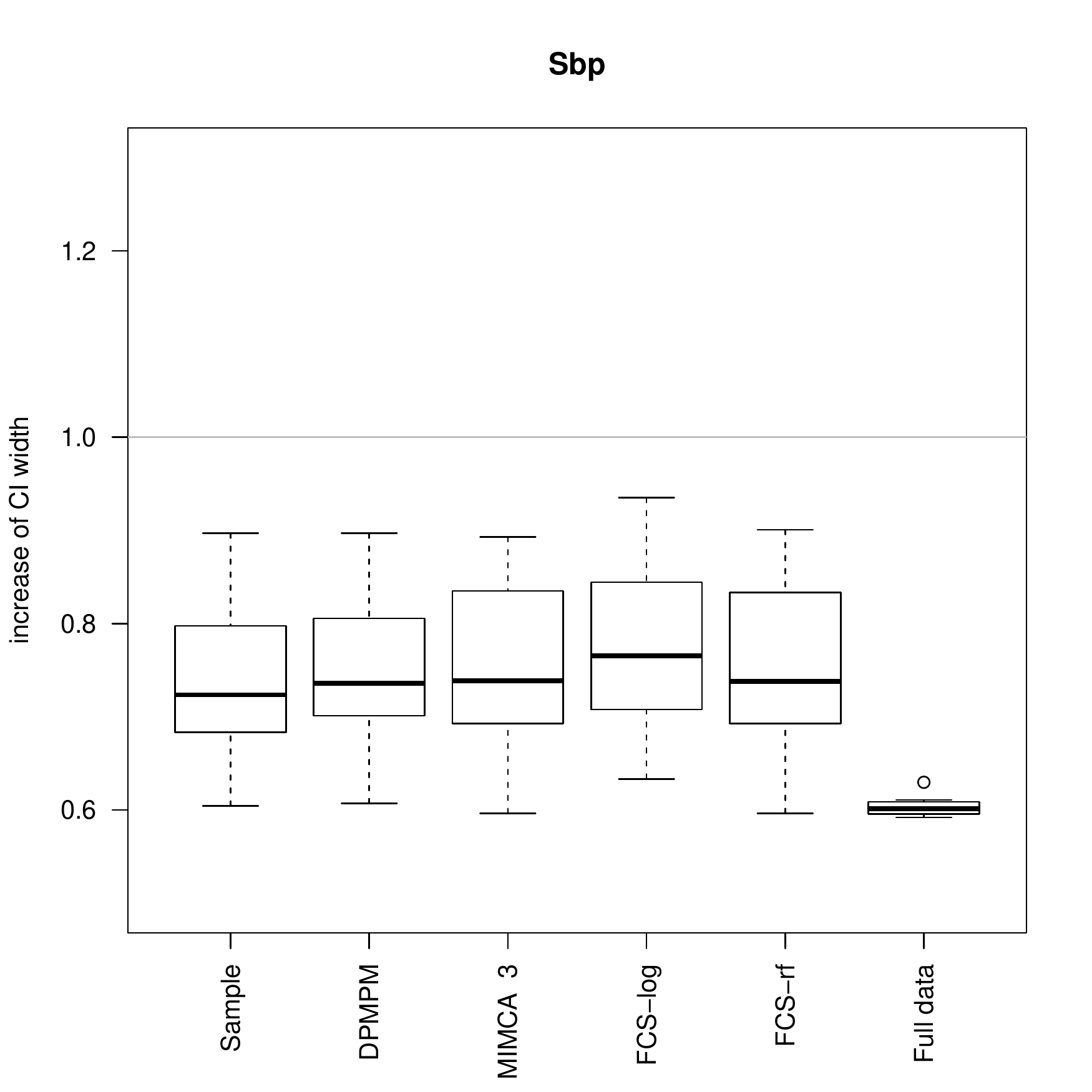}
\includegraphics[scale=.3]{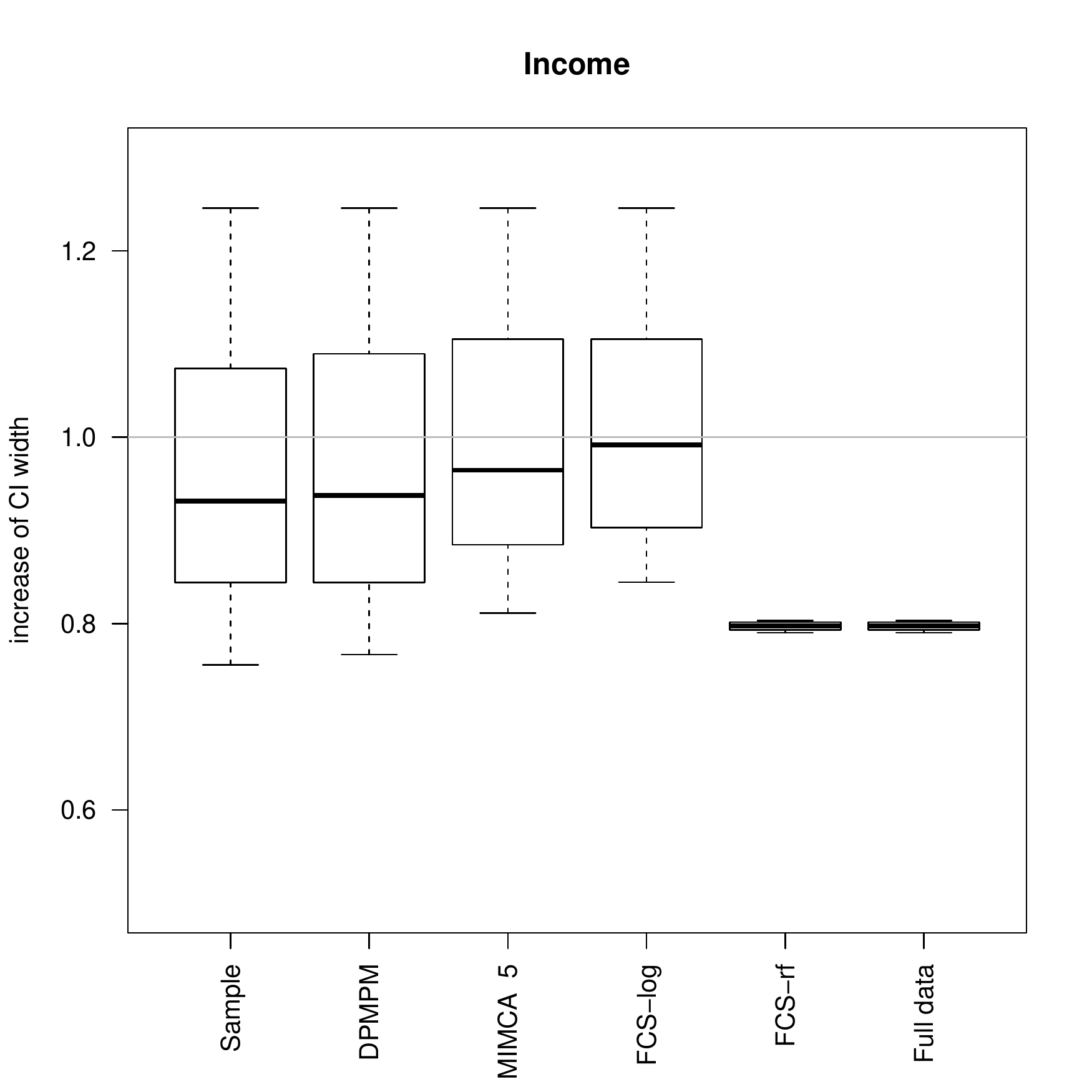}
\includegraphics[scale=.3]{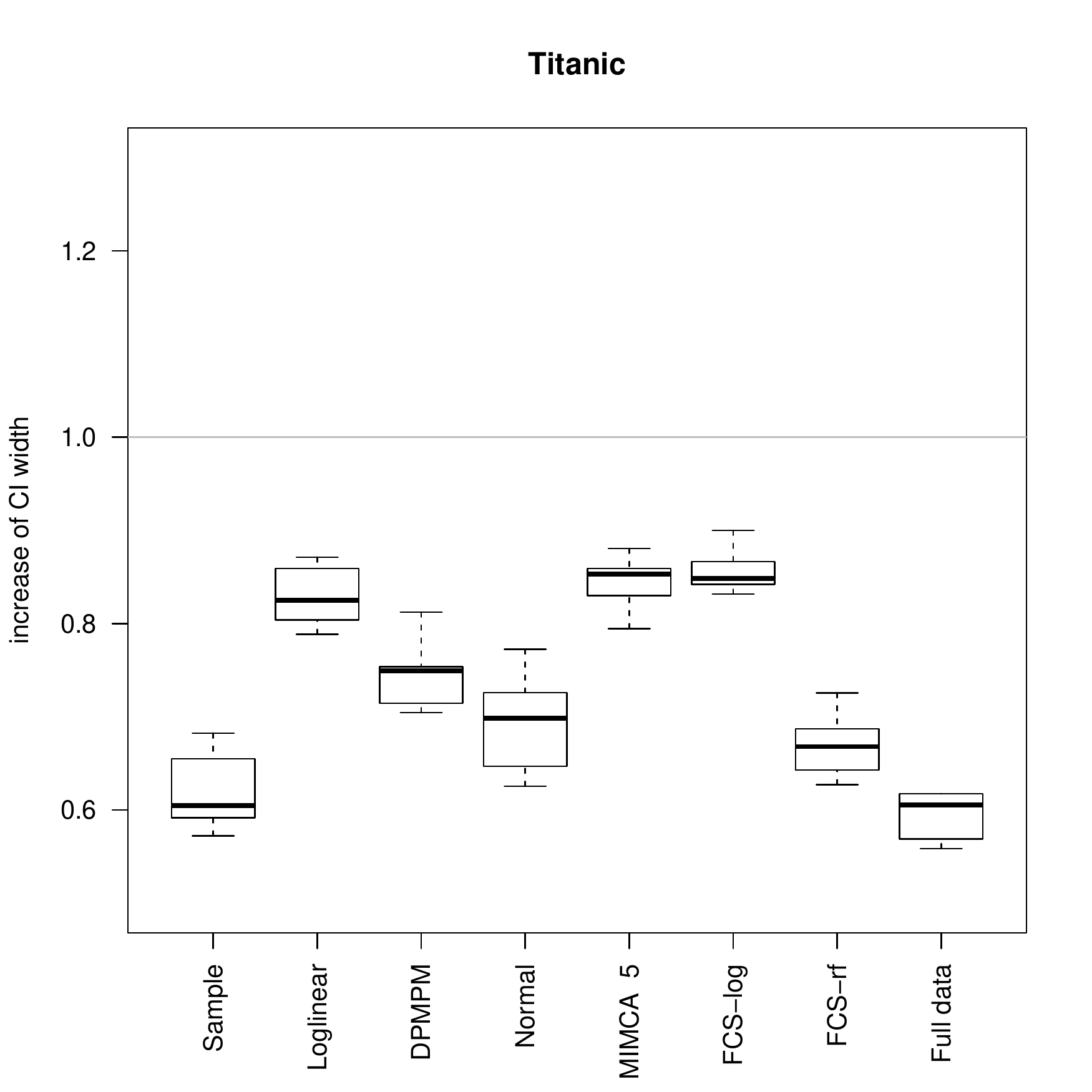}
\includegraphics[scale=.3]{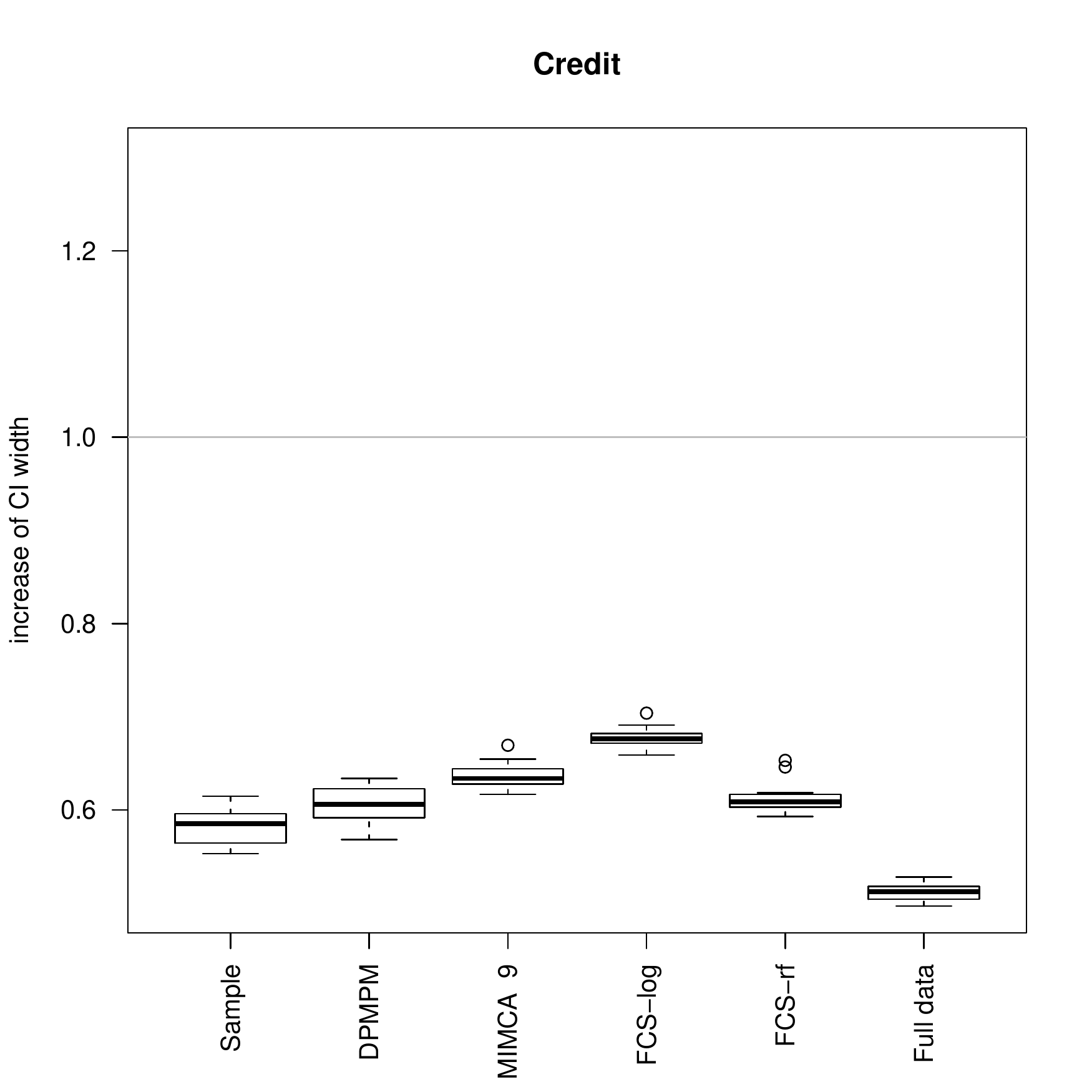}
\caption{Distribution of the median of the confidence interval for the several quantities of interest for several methods (Sample, Loglinear model, DPMPM, Normal distribution, MIMCA, FCS using logistic regressions, FCS using random forests, Full data) for different data sets (Saheart, Galetas, Sbp, Income, Titanic, Credit). One point represents the median of the confidence interval observed for one coefficient divided by the one obtained by Listwise deletion. The horizontal dashed line corresponds to a ratio of 1. Points over this line corresponds to confidence interval higher than the one obtain by listwise deletion.\label{figci}}
\end{center}
\end{figure}
\begin{figure}[h!]
\begin{center}
\includegraphics[scale=.3]{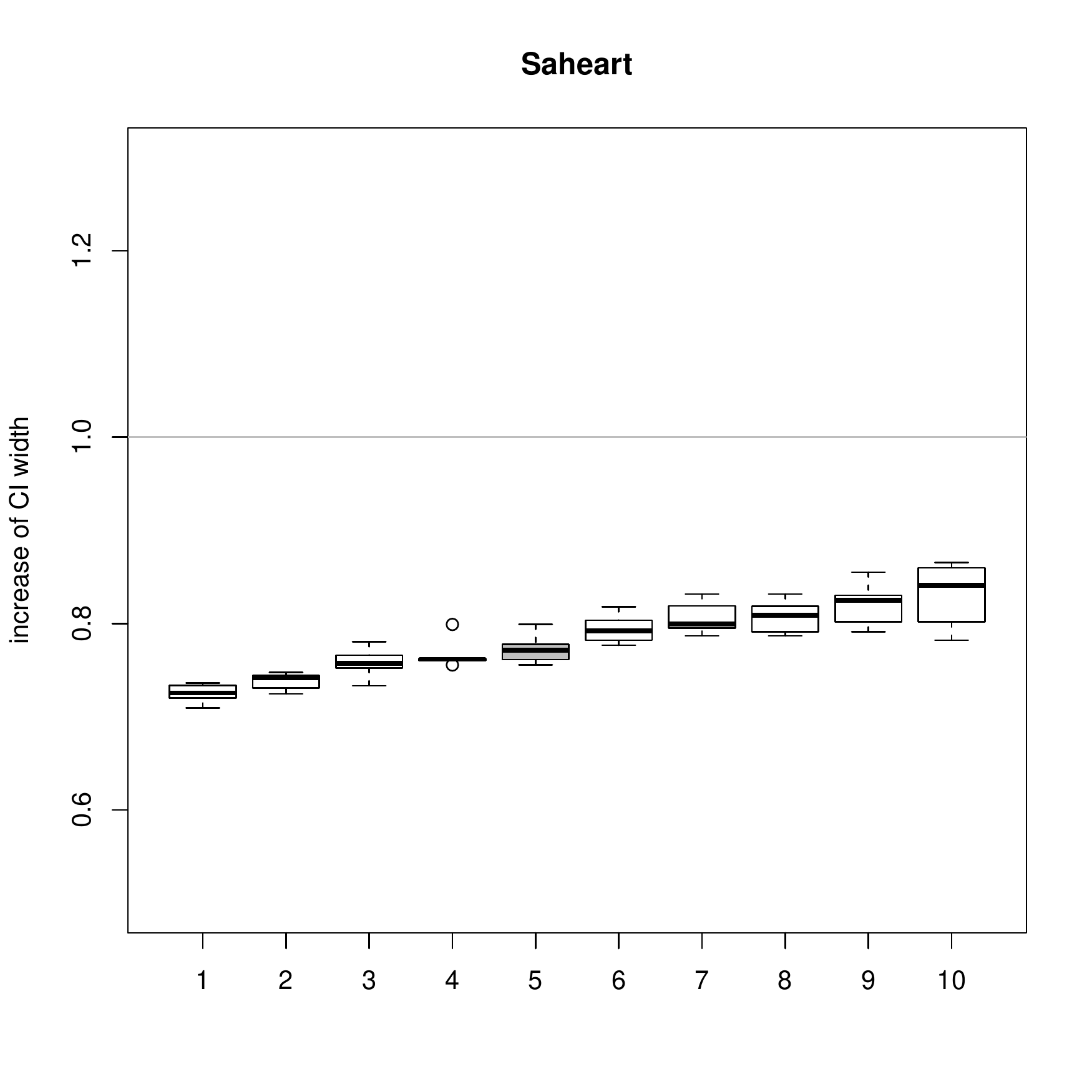}
\includegraphics[scale=.3]{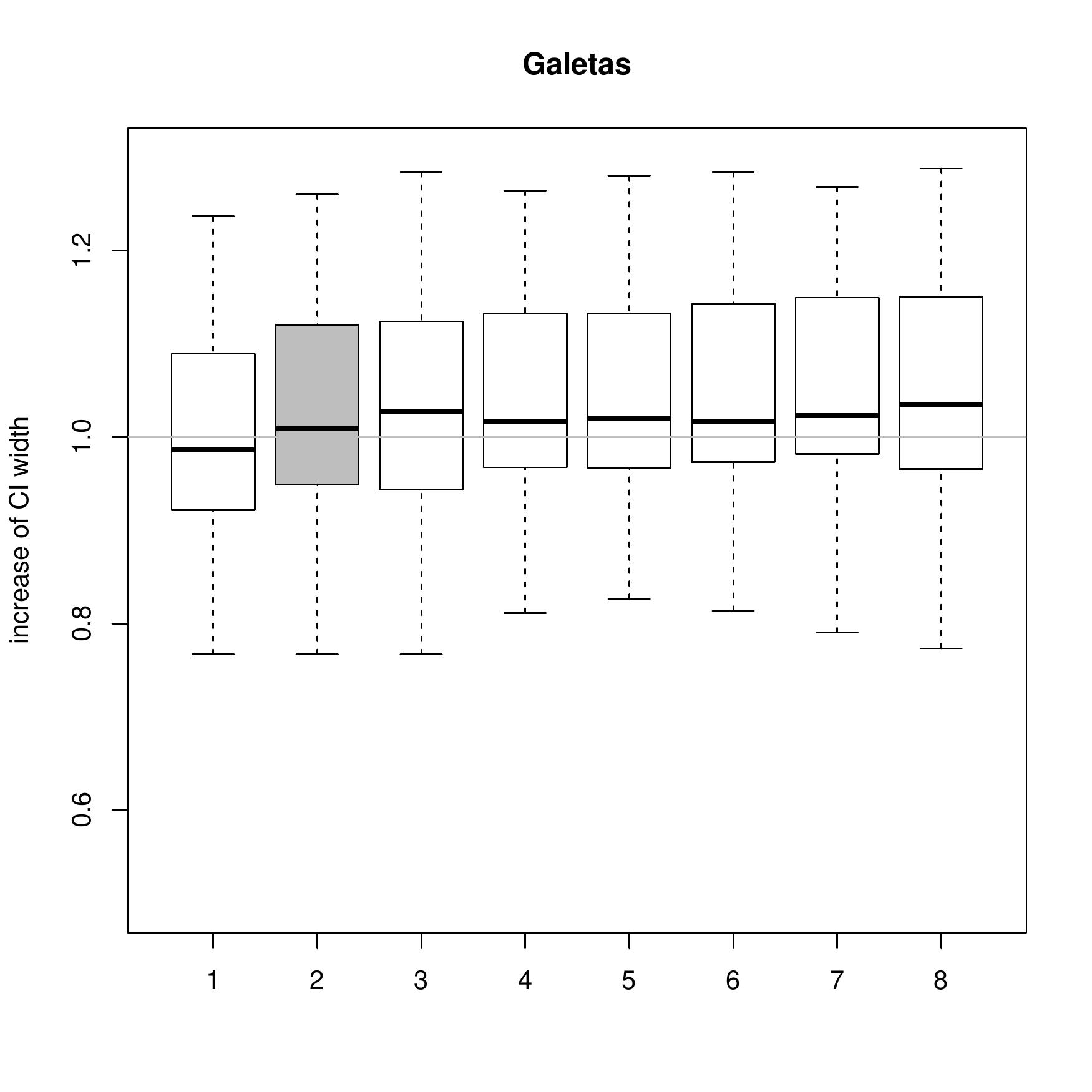}
\includegraphics[scale=.3]{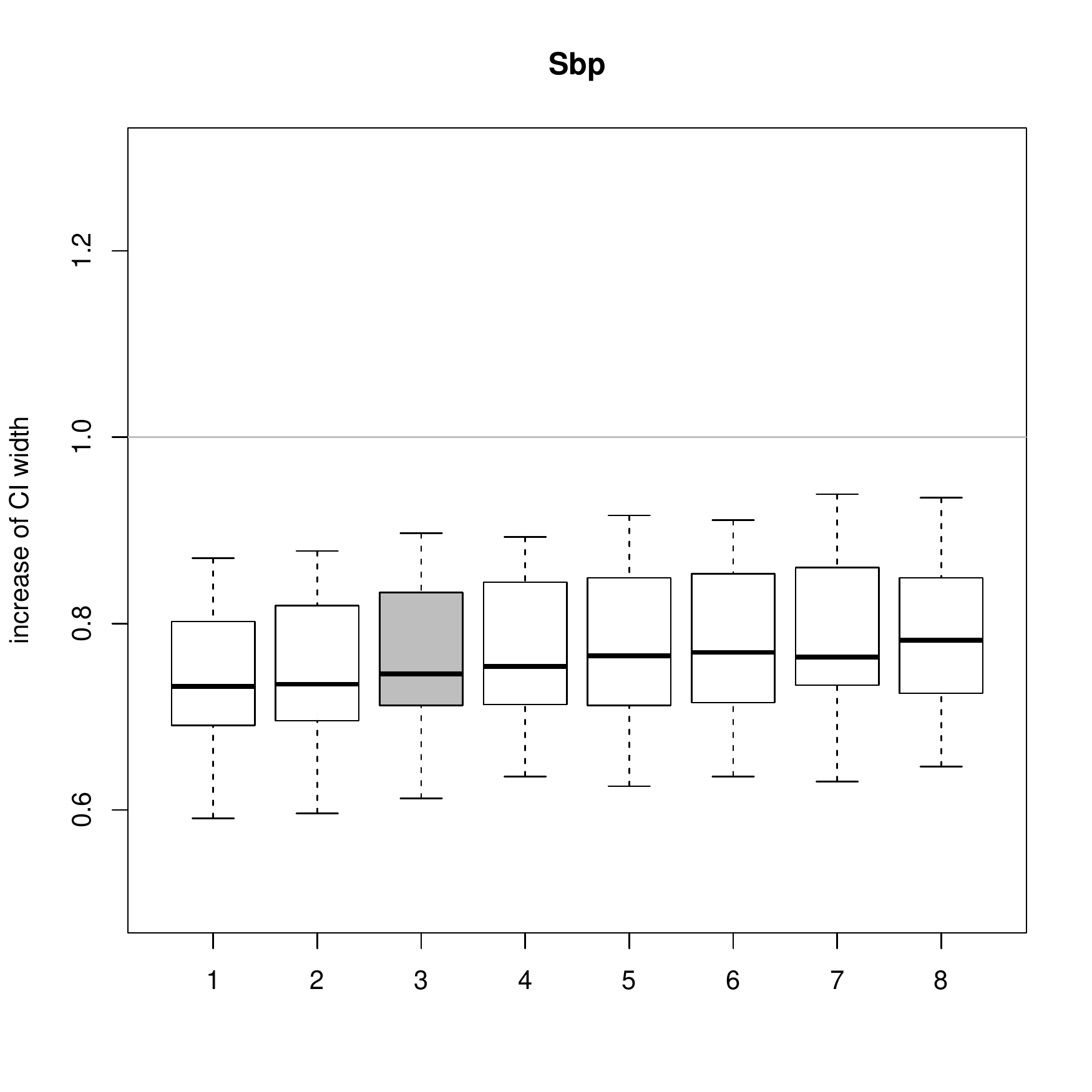}
\includegraphics[scale=.3]{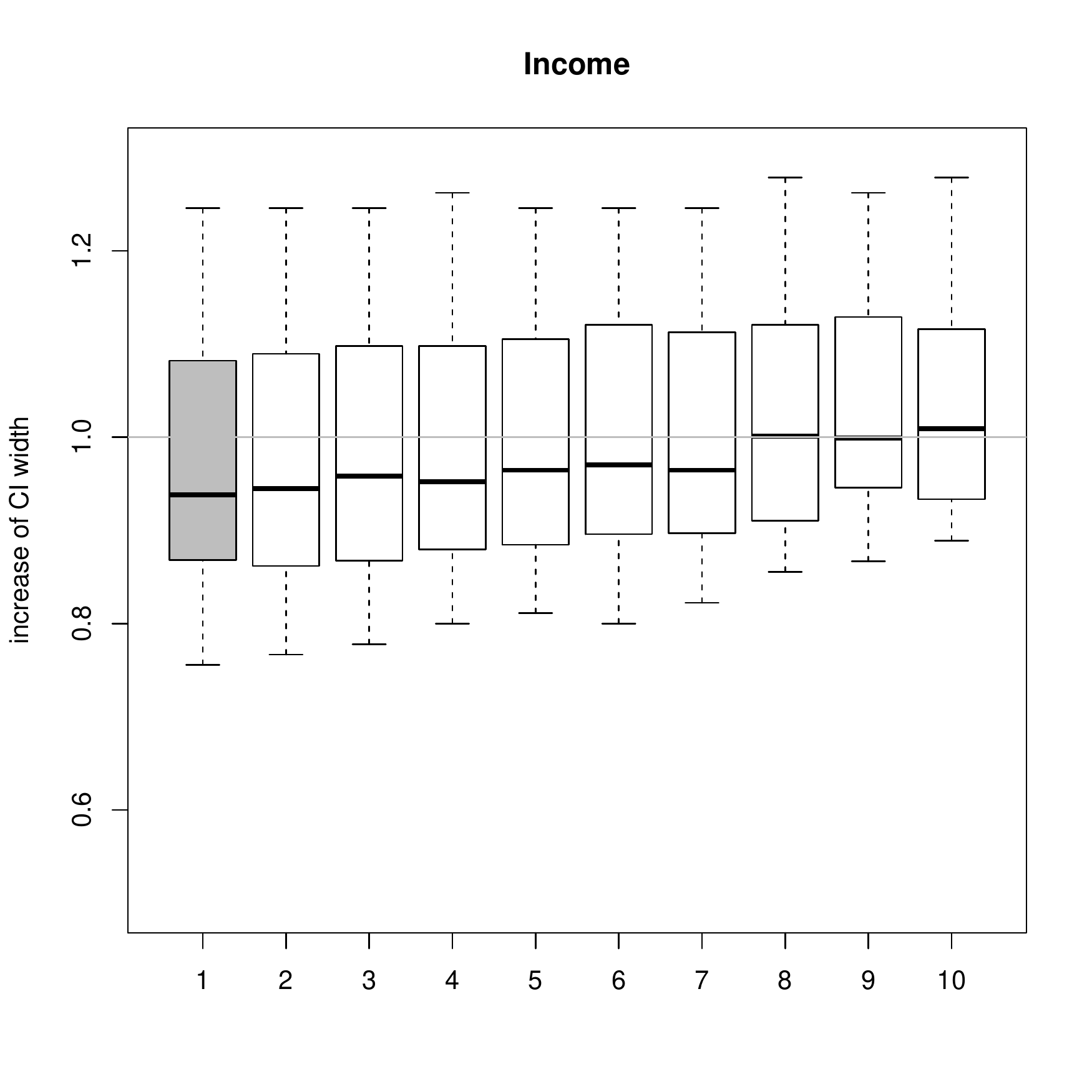}
\includegraphics[scale=.3]{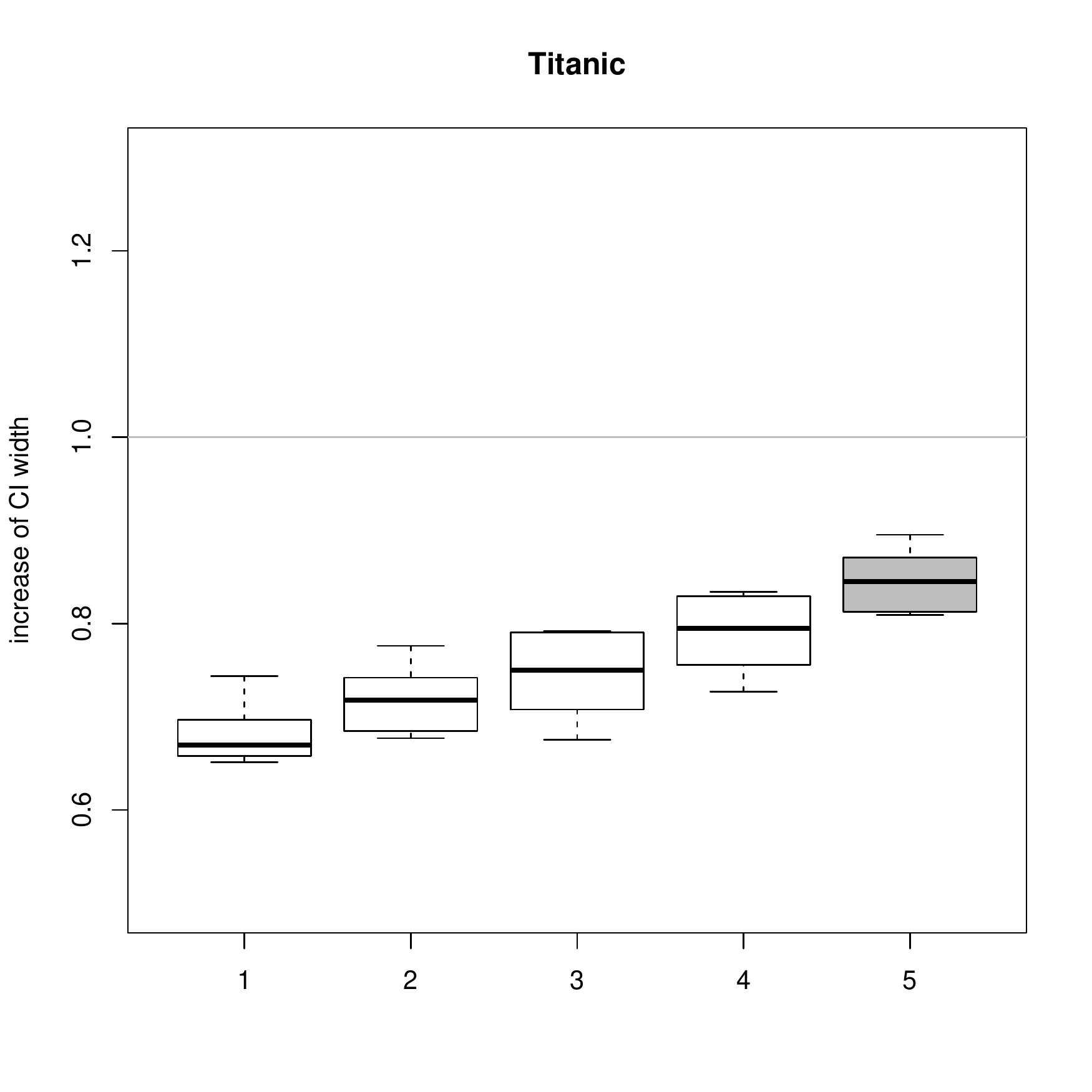}
\includegraphics[scale=.3]{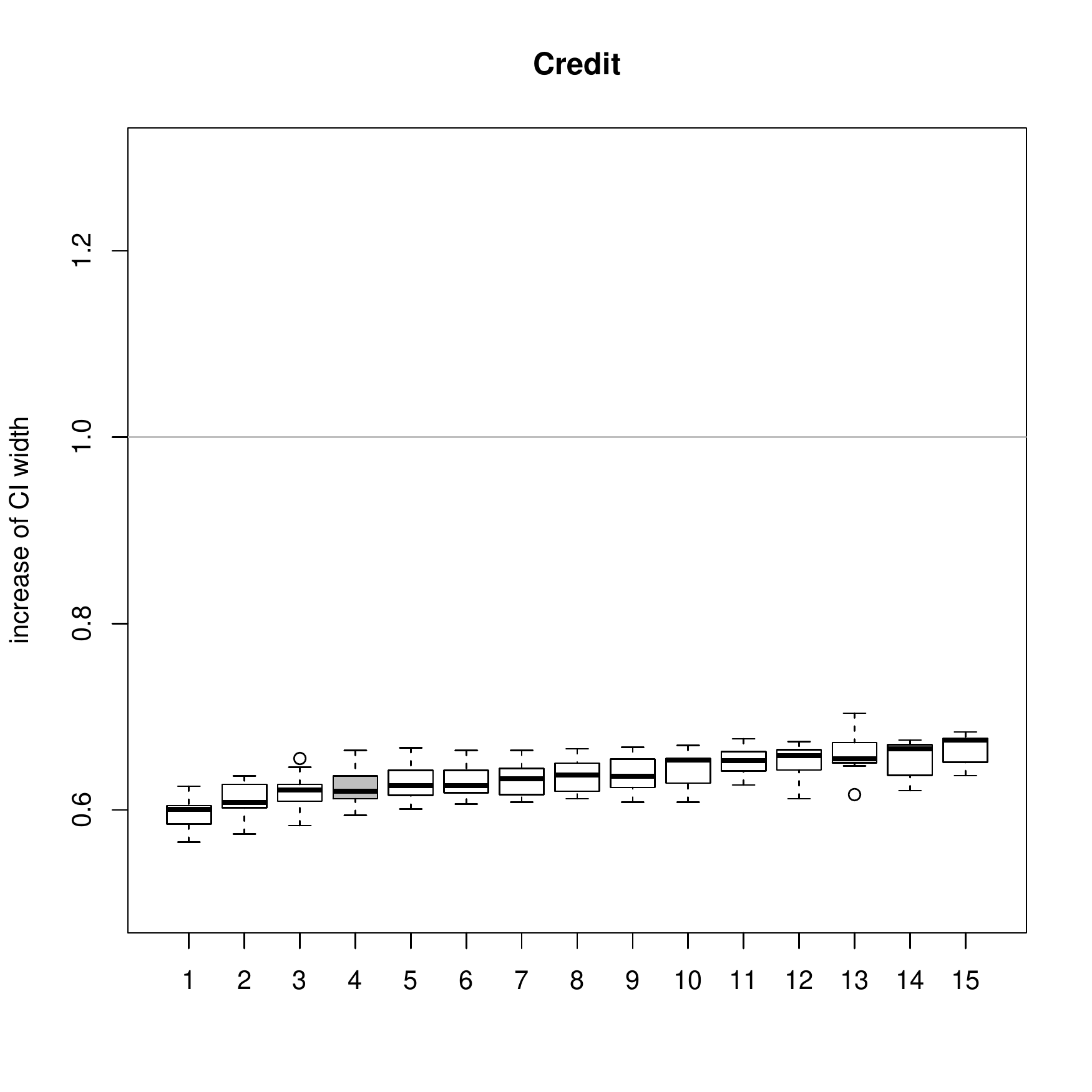}
\caption{Distribution of the median of the confidence interval for the several quantities of interest for the MIMCA algorithm for several numbers of dimensions for different data sets (Saheart, Galetas, Sbp, Income, Titanic, Credit). One point represents the median of the confidence interval observed for one coefficient divided by the one obtained by Listwise deletion. The horizontal dashed line corresponds to a ratio of 1. Points over this line corresponds to confidence interval higher than the one obtain by listwise deletion. The results for the number of dimensions provided by cross-validation are in grey.\label{fignbaxeci}}
\end{center}
\end{figure}
\begin{figure}[h!]
\begin{center}
\includegraphics[scale=.3]{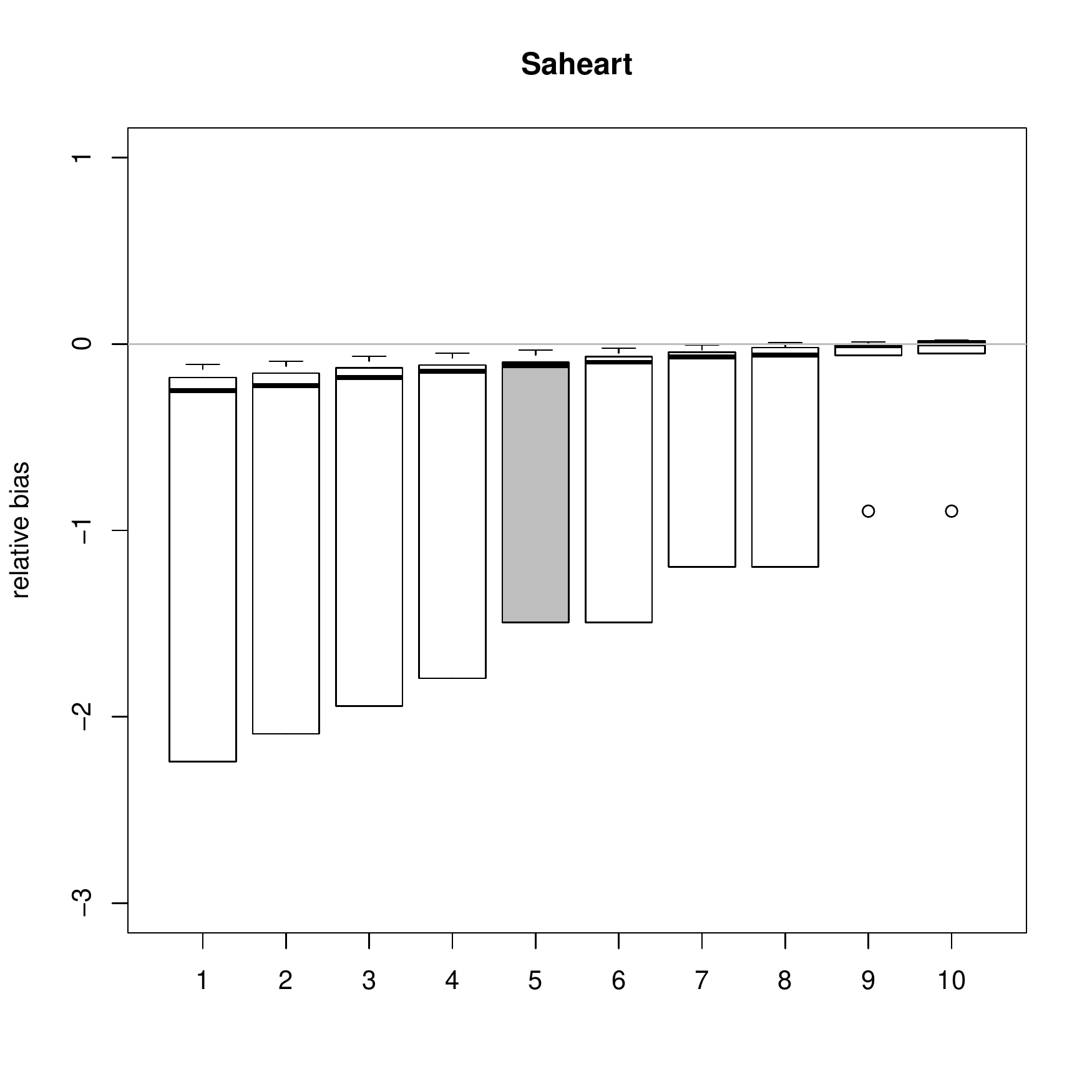}
\includegraphics[scale=.3]{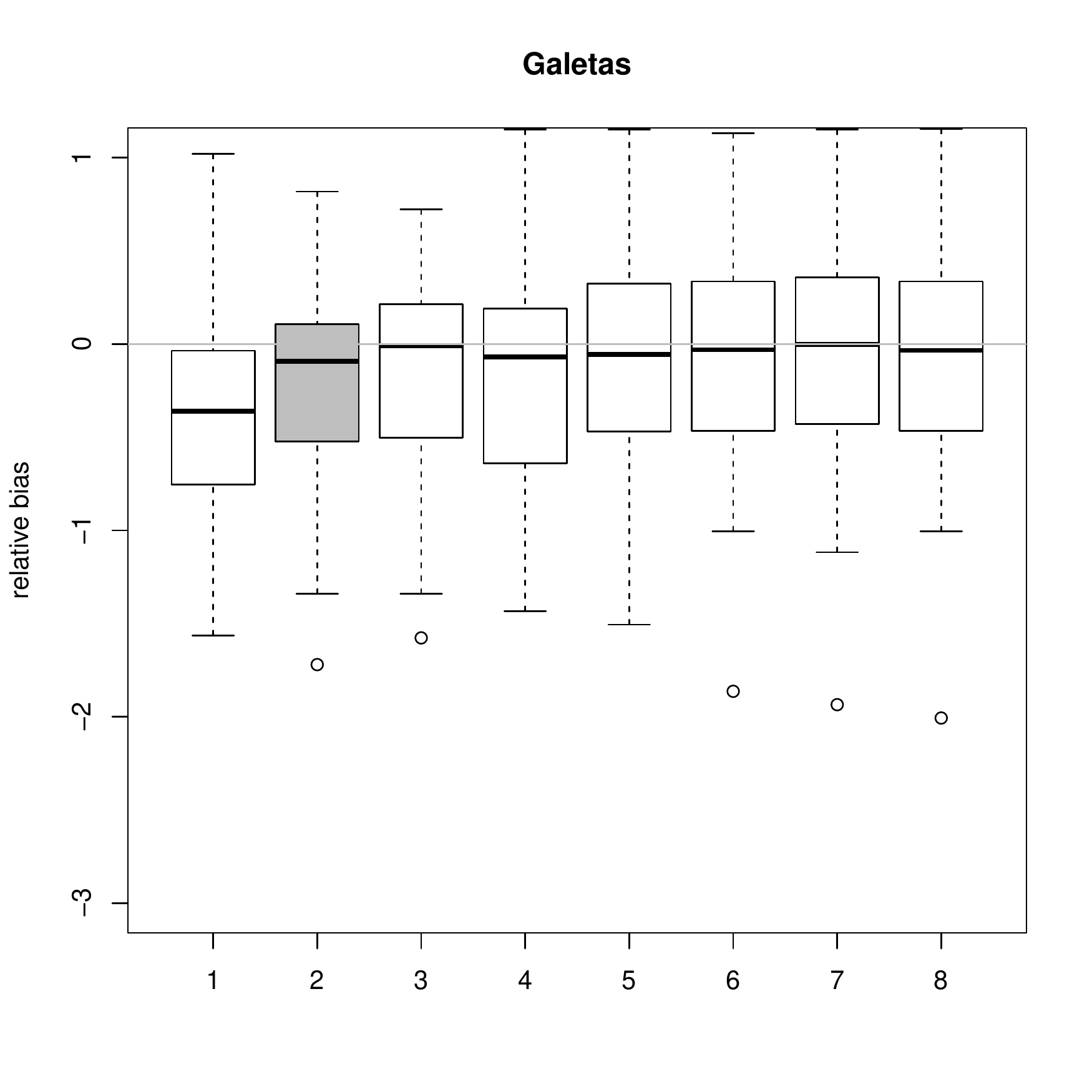}
\includegraphics[scale=.3]{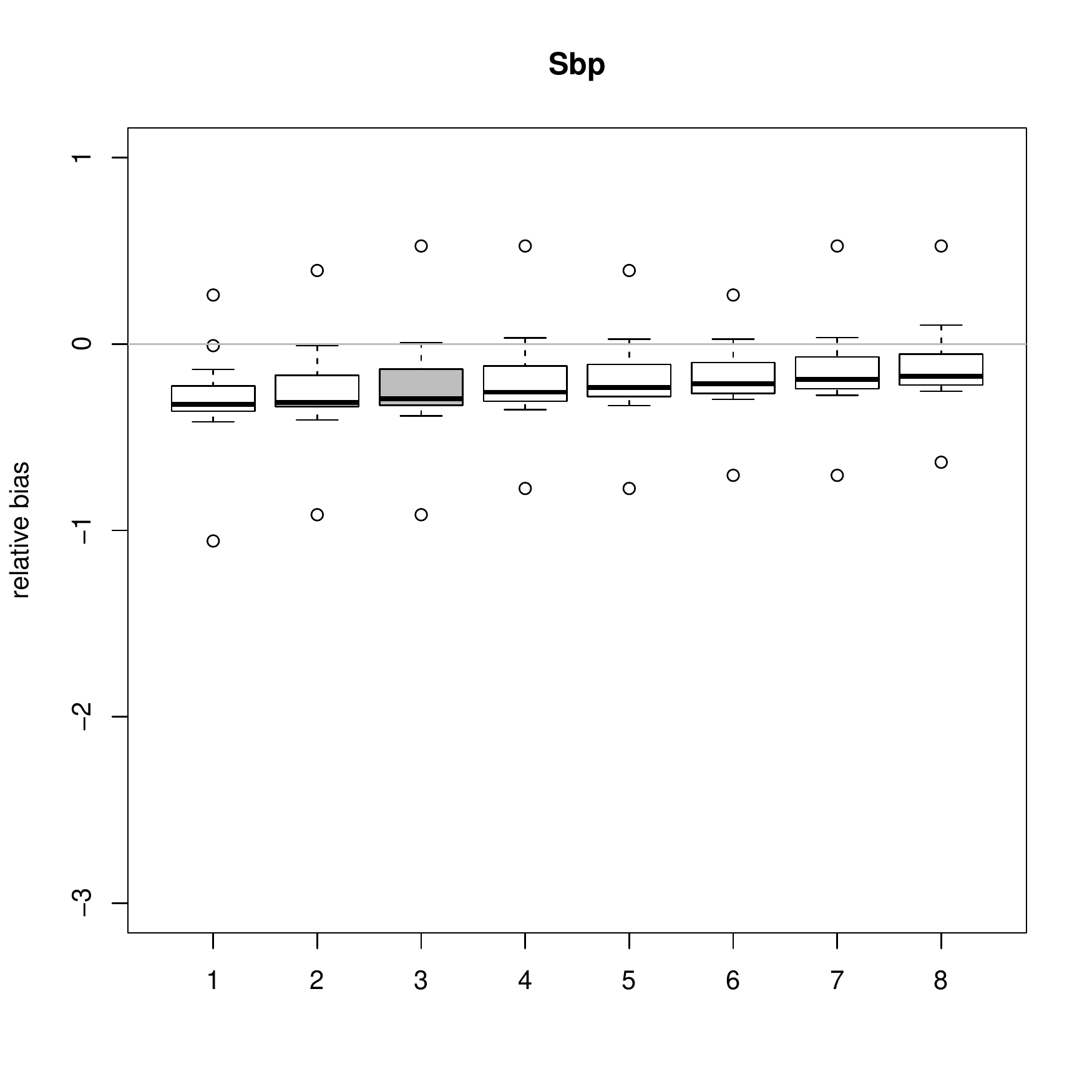}
\includegraphics[scale=.3]{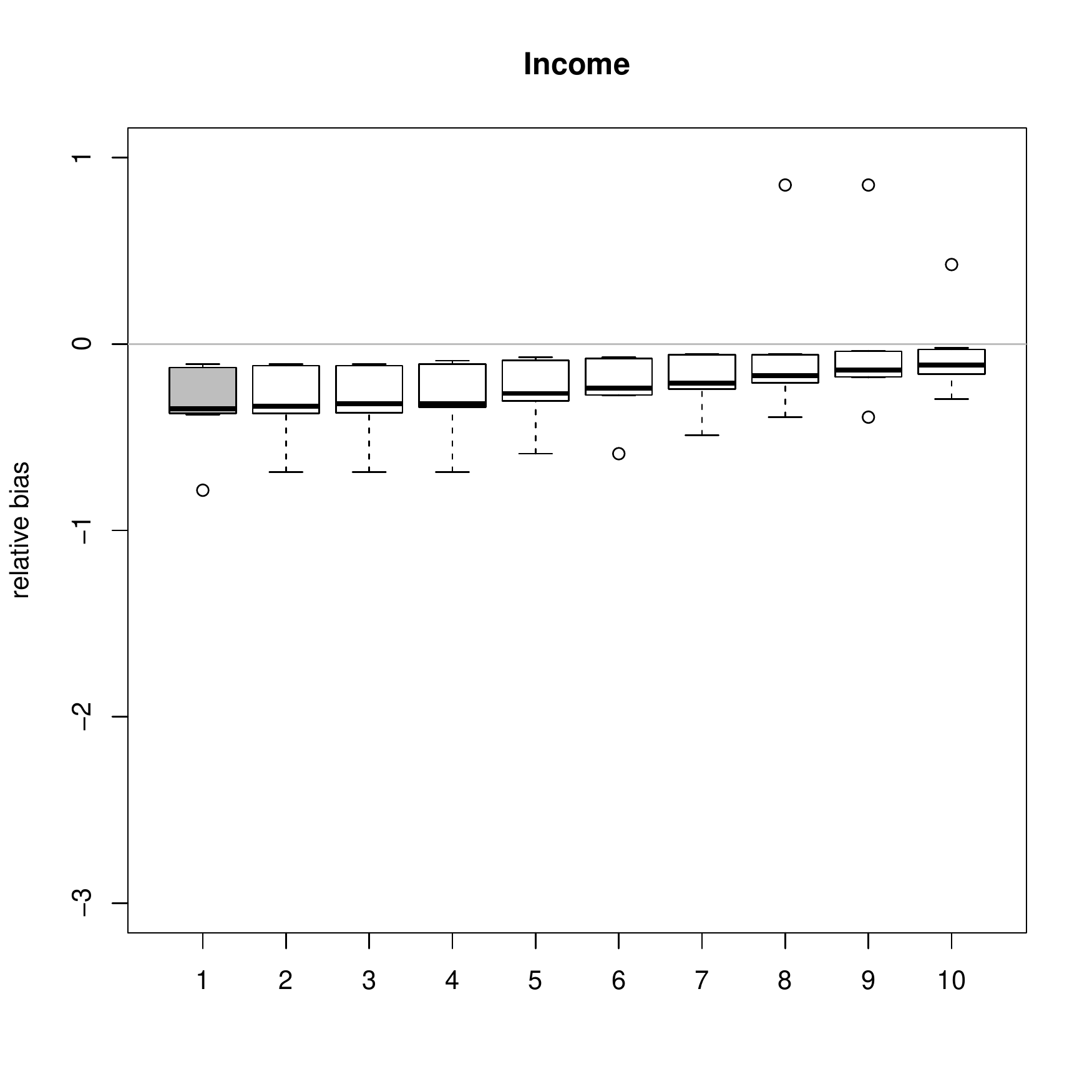}
\includegraphics[scale=.3]{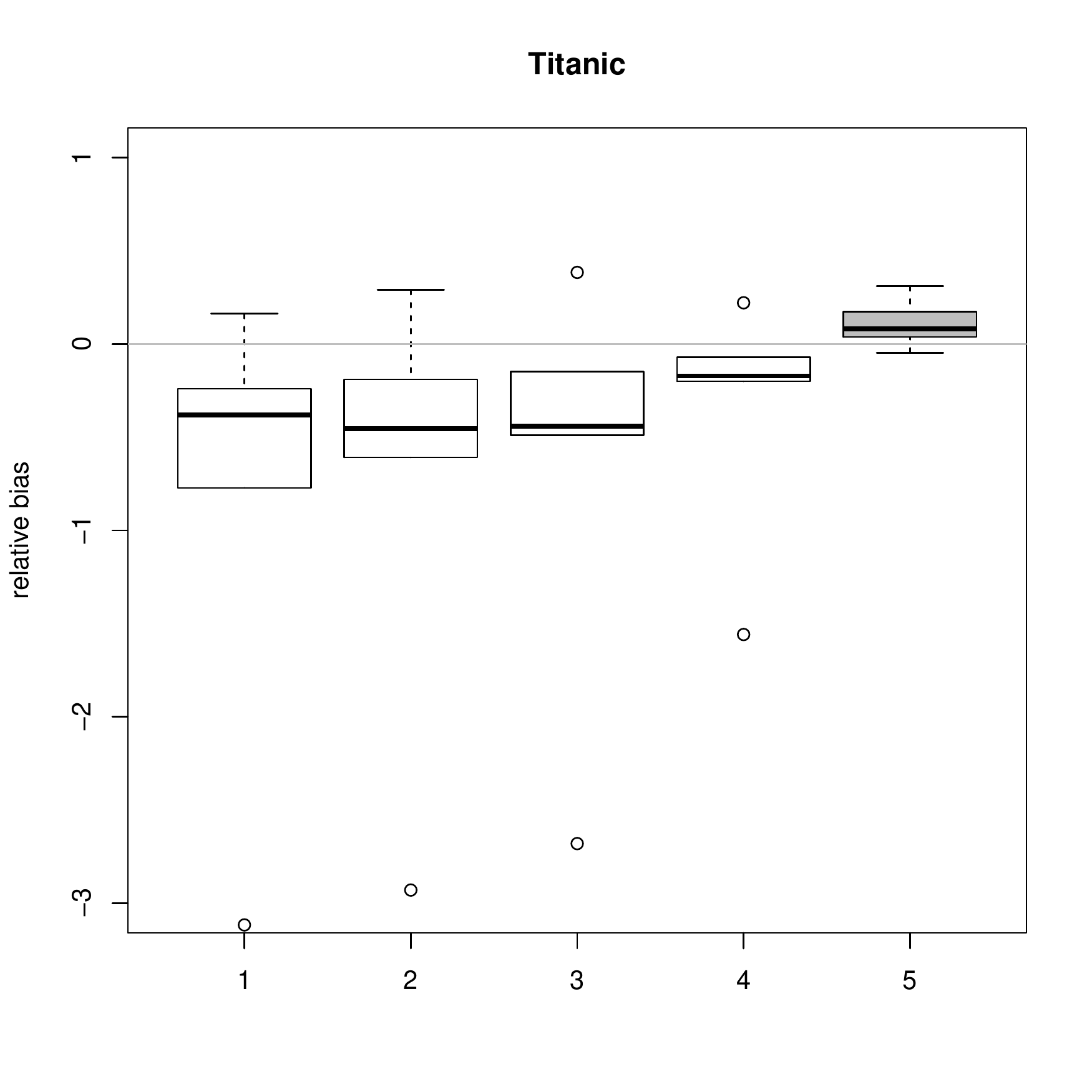}
\includegraphics[scale=.3]{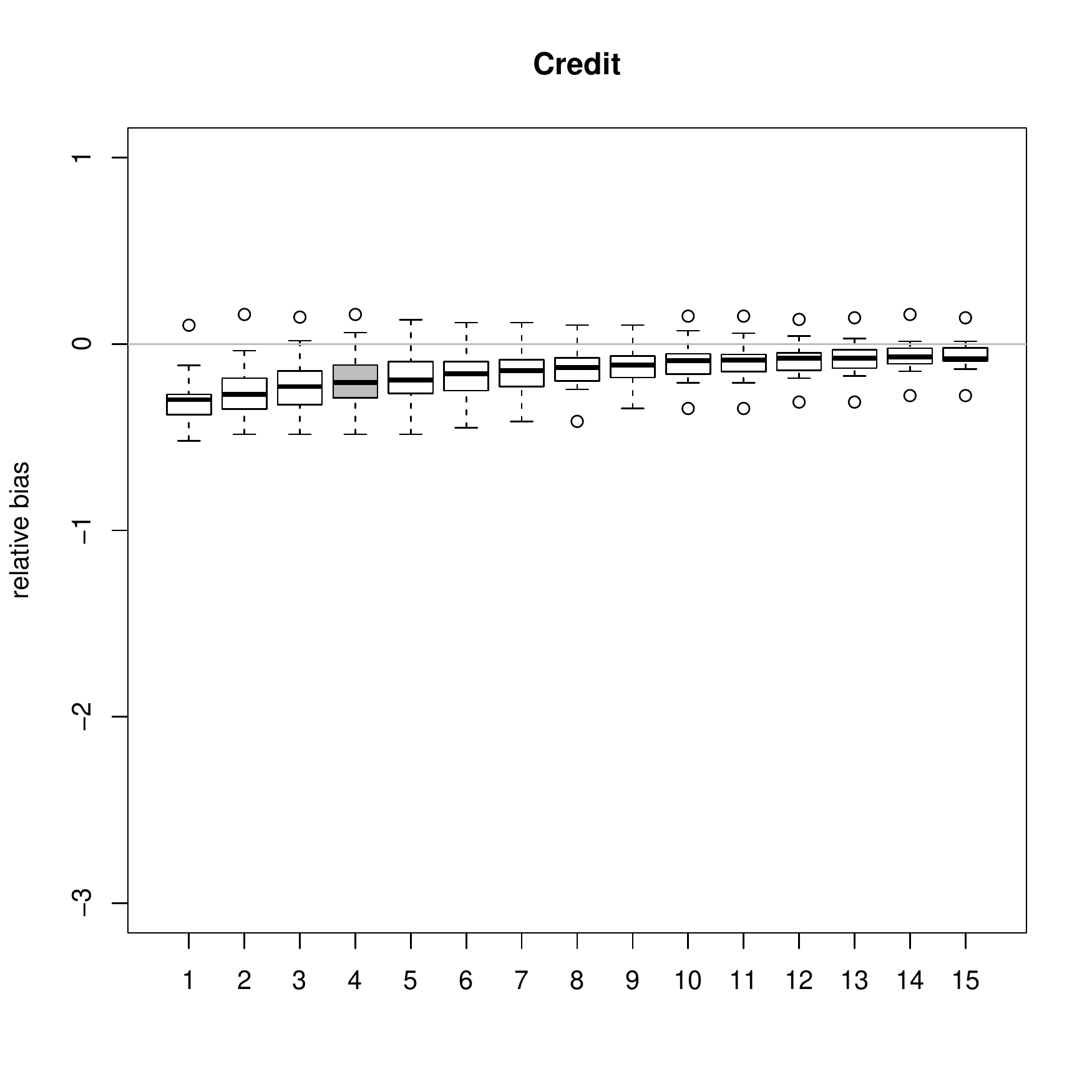}
\caption{Distribution of the relative bias (bias divided by the true value) over the several quantities of interest for the MIMCA algorithm for several numbers of dimensions for different data sets (Saheart, Galetas, Sbp, Income, Titanic, Credit). One point represents the relative bias observed for one coefficient. The results for the number of dimensions provided by cross-validation are in grey. \label{fignbaxebias}}
\end{center}
\end{figure}
\end{document}